\documentclass{jfm}

\usepackage{graphicx}
\usepackage{newtxtext}
\usepackage{newtxmath}
\usepackage{natbib}
\usepackage{hyperref}
\usepackage{caption}
\usepackage{subcaption}

\DeclareCaptionFormat{tagleft}{%
  {#1#2}{#3}}

\hypersetup{
    colorlinks = true,
    urlcolor   = blue,
    citecolor  = black,
}

\newcommand{\RomanNumeralCaps}[1]
\linenumbers

\title{Harmonic structure and parametric prediction of instantaneous power in oscillating foils}

\author{Junjie Shen\aff{1}, 
  \corresp{\email{jshen89@wisc.edu}},
  Jennifer A. Franck\aff{1}
  \corresp{\email{jafranck@wisc.edu}}
  }

\affiliation{\aff{1}Department of Mechanical Engineering,
  University of Wisconsin--Madison, Madison, WI 53706, USA}

\begin{document}
\maketitle

\begin{abstract}
A closed-form model is developed to predict the instantaneous lift, pitching moment, and power of an oscillating-foil energy harvester directly from its kinematic input. Such foils extract renewable power from moving water through sinusoidal heaving and pitching. However, due to their large-amplitude motion and vortex-dominated dynamics, classical unsteady aerodynamics cannot accurately predict the time-dependent force and power generation. To generate the model, 405 two-dimensional simulations are computed which vary heave amplitude, pitch amplitude, and frequency, representing the energy harvesting kinematic parameter space. The lift and pitching moment are decomposed into odd Fourier harmonics, and each Fourier coefficient is regressed on sixteen physics-based polynomial terms derived from the kinematic, power, and energy scalings. The regression is performed in two Ridge steps, one for attribution and one for the deployable predictor. The attribution shows that the first harmonic, which contains the mean power, depends on the motion amplitudes. The higher harmonics respond nonlinearly to the frequency, which determines the number and timing of the vortex shedding. The model recovers the cycle-averaged power with a leave-one-out $R^2$ of $0.98$, and reproduces the instantaneous lift with a median peak error of $9\%$ and a median phase error below $1\%$ of the period. The harmonic order of the response counts the discrete vortex events in each stroke, as confirmed by the wake visualization. The model's analytical expression for time-dependent force and power can be utilized as a fast and accurate alternative to high-fidelity simulations, specifically targeting device design and array optimization of oscillating foils.

\end{abstract}



\section{Introduction}
\label{sec:intro}

This paper develops a closed-form algebraic model to predict the instantaneous lift, pitching moment and power of an oscillating-foil energy harvester from its kinematic input. Oscillating foils have been researched as energy harvesters that operate periodically in both heave and pitch.
Particularly suited to marine energy applications, oscillating foils can operate in tandem, or in arrays, spanning rectangular cross-sections of a channel efficiently to produce clean and renewable power from moving water. Although there is a substantial body of fundamental fluid mechanics research on oscillating foils from flapping flight \citep{Sane2003, ChinLentink2016} and propulsion \citep{Triantafyllouetal2000, Smits2019}, the energy-harvesting mode is distinct in that it generates a drag wake and operates most efficiently at high pitch and heave amplitudes~\citep{KinseyDumas2008, RibeiroFranck2020}. These flow features violate most quasi-steady and linearized aerodynamic theories, rendering traditional analytical models for time-dependent forces unreliable. As an alternative approach, this research proposes a data-driven parametric prediction of the instantaneous forces. This is accomplished by decomposing the periodic force and moment into harmonics and reconstructing each harmonic with physics-derived features, revealing an analytic expression for the time-dependent power.


In the energy harvesting mode the oscillating foil is a turbine, producing net drag and extracting kinetic energy from the flow as first demonstrated by \citet{McKinneyDeLaurier1981}. 
Oscillating foils for energy harvesting operate at lower frequencies and higher amplitudes of heave and pitch than propulsive foils \citep{Youngetal2014, XiaoZhu2014, Wuetal2020, Triantafyllouetal1991, Andersonetal1998}, generating power through the lift force and heave velocity. 
The optimal kinematic regime is when the sinusoidal pitch and heave are $90^\circ$ out of phase, such that the foil has a positive angle of attack and high lift aligned with a positive heave velocity. A rotation at the top of the heave stroke then sets up the reverse motion with a negative angle of attack on the downstroke, initiating the second power stroke. For small pitch and heave amplitudes the boundary layer often remains attached throughout this process, and the time-dependent lift and moment are roughly sinusoidal, following the sinusoidal kinematic input. However, to maximize energy generation, high amplitudes and high effective angles of attack are required, in which the flow separates at the leading edge producing a leading-edge vortex (LEV), and sometimes an accompanying trailing-edge vortex (TEV). These vortices produce transient peaks in the force, moment and power \citep{RibeiroFranck2020, SialaLiburdy2019}. The number, timing, and strength of these peaks shift with only slight changes to the kinematics, so the instantaneous force at otherwise similar operating points can differ substantially. These changes in shape have been widely reported \citep{Kimetal2017, Ribeiroetal2021}, but their quantitative relationship with the input kinematic parameters has not been established.

The leading- and trailing-edge vortices that produce these peaks are shed in a periodic pattern and convect downstream into a coherent wake that has a direct correspondence with the oscillation kinematics \citep{RibeiroFranck2023}. The link between wake structure and kinematics has also been investigated for propulsive foils, beginning with experimental maps of the wake pattern over the kinematic space \citep{Schnipperetal2009}. Machine-learning models have since inverted this map, identifying the wake type or the underlying motion from the flow field \citep{Colvertetal2018, Lietal2020, Calvetetal2021, Jietal2023}. This correspondence is important to understanding how multiple oscillating foils can interact with one another in arrays or multi-bodied systems \citep{KinseyDumas2012, Ribeiroetal2021, HandyCardenasetal2025}. In developing such interaction models, it is paramount to have a prediction of the time-dependent forces and power generation such that the timing of the vortex can properly interact with the transient energy harvesting process \citep{RibeiroFranck2024}.


Predicting performance or forces directly from the kinematics has progressed furthest for propulsive oscillating foils. For example, \citet{Floryanetal2017} derived scaling laws that relate the heave and pitch kinematics to the cycle-averaged thrust and power. The model contains added-mass and circulatory contributions with the Strouhal number as the leading parameter. \citet{VanBurenetal2019} extended these scalings to foils that heave and pitch together, whereas \citet{MooredQuinn2019} developed inviscid scaling laws for a self-propelled pitching foil. These laws are functions of input kinematics alone and are derived from physical mechanisms, but they are formulated to predict the mean thrust and efficiency of propulsive foils rather than the instantaneous forces.

Energy-harvesting foils are governed by different scaling laws due to the shift from thrust-producing to energy-generation physics. Classical unsteady aerodynamics gives closed-form instantaneous forces \citep{Theodorsen1935, Garrick1936}, but it assumes small-amplitude motion and attached flow, conditions that the large-amplitude, separated regime of harvesting violates. Within the energy harvesting literature, a common parameterization of kinematics is the quarter-period effective angle of attack $\alpha_{T/4}$ \citep{KinseyDumas2008}, which can collapse the mean power and efficiency over a wide kinematic range \citep{Kimetal2017} and can be used to classify kinematics into shear-layer, LEV, and LEV+TEV regimes \citep{Ribeiroetal2021, Leeetal2022}. However, it can be shown that kinematics with the same value of $\alpha_{T/4}$ can produce surprisingly different instantaneous forces, moment and power, and different wake topologies \citep{RibeiroFranck2023}. Even the optimal regimes require a second parameter, the rate of change of the effective angle of attack \citep{KinseyDumas2014}.

Parameterizing the unsteady forces from the foil kinematics has recently been attempted for flapping flight. For example, \citet{Nakataetal2015} built a quasi-steady model with force coefficients fitted by least squares to high-fidelity simulations. \citet{Caladoetal2023} combined Ridge regression with Gaussian processes to model the aerodynamic forces under different hovering kinematics, whereas \citet{Kamimizuetal2025} used sparse identification to refine an existing quasi-steady model to predict the instantaneous force from the wing kinematic variables. 


To date, there is no parametric model that predicts the time-dependent power directly from the kinematic input for oscillating foils. Such a prediction model would enable rapid evaluation of the design space, and give insight into the unsteady vortex dynamics by means of transient peaks in lift and moment cycles. Furthermore, such a model is necessary when considering arrays of multiple oscillating foils, as the interaction between the wake and subsequent oscillating foils is fundamentally time-dependent, and intractable with time-averaged models. 

To address this gap, this research derives a time-dependent prediction model for the lift and moment as functions of the kinematic variables. The unsteady forces are decomposed into Fourier harmonics, and each Fourier coefficient is predicted from the kinematics by Ridge regression on physics-derived features. The specific contributions of this paper are (i) a closed-form model that maps the three kinematic parameters (frequency, heave amplitude, and pitch amplitude) to the instantaneous power, (ii) a harmonic-resolved attribution that identifies the kinematic groups behind each Fourier coefficient, and (iii) a vortex-event interpretation that links the high-order harmonics ($k=3,5$) to discrete LEV and LEV+TEV shedding.

The paper is organized as follows. Section~\ref{sec:methods} describes the foil kinematics and the computational methods. The prediction model with Fourier decomposition and feature design is set up in \S\ref{sec:modeling}. The results follow in \S\ref{sec:results}, where the Ridge attribution of each harmonic, the closed-form predictor, and its error, instantaneous-response, and peak-phase validation are examined. We then interpret the high-order harmonics in terms of vortex-shedding events and compare the lift with Theodorsen's model (\S\ref{sec:shedding}), before summarizing the conclusions in \S\ref{sec:discussion}.

\section{Simulation data and methods}
\label{sec:methods}

Building the model requires the instantaneous forces and moments across the kinematic space of energy harvesting oscillating foils, which will serve as targets of the prediction model. This section describes the computational methods and kinematics (input parameter space) of the simulations performed. 

\subsection{Computational methods}
\label{sec:cfd}

The flow is governed by the two-dimensional (2D) incompressible Navier--Stokes equations at $\Rey = U_\infty c / \nu = 1000$, where $U_\infty$ is the freestream velocity, $c$ the chord length, and $\nu$ the kinematic viscosity. The equations are solved using the finite-volume pressure-implicit split-operator (PISO) method implemented in OpenFOAM \citep{Welleretal1998}. A low Reynolds number enables a broad parametric sweep within reasonable computational cost, and prior work has shown that the power generation and wake characteristics are qualitatively similar across $\Rey = 1000$--$50{,}000$ \citep{RibeiroFranck2020, Ribeiroetal2021}.

The governing equations, solver settings, computational domain and mesh follow the setup of \citet{RibeiroFranck2023}, who performed a mesh convergence study with eight meshes of varying resolution. An unstructured mesh of $1.64 \times 10^5$ nodes is implemented, with 150 nodes along the foil surface, a near-foil resolution of $\Delta x / c = 0.013$, and a wake resolution of $\Delta x / c = 0.05$. The foil motion is imposed through a dynamic mesh solver that updates all node positions at each time step. The domain extends $106c$ in the streamwise direction and $100c$ in the cross-stream direction, with the foil located $50c$ from the inlet. A no-slip condition is applied on the foil surface, with uniform flow at the inlet and zero-gradient outflow on the remaining boundaries.

Each simulation computes four oscillation cycles. The forces converge to a periodic state after three cycles, and the fourth cycle is extracted for analysis. The forces and moment are sampled at 256 points per oscillation cycle, and the velocity field is saved at 64 uniform time instances per cycle.

\subsection{Oscillating foil kinematics, dynamics and power generation}
\label{sec:power}

\begin{figure}
    \centering
    \begin{subfigure}[t]{0.60\textwidth}
      \caption{}\label{fig:1a}
      \centering
      \includegraphics[width=1.00\linewidth]{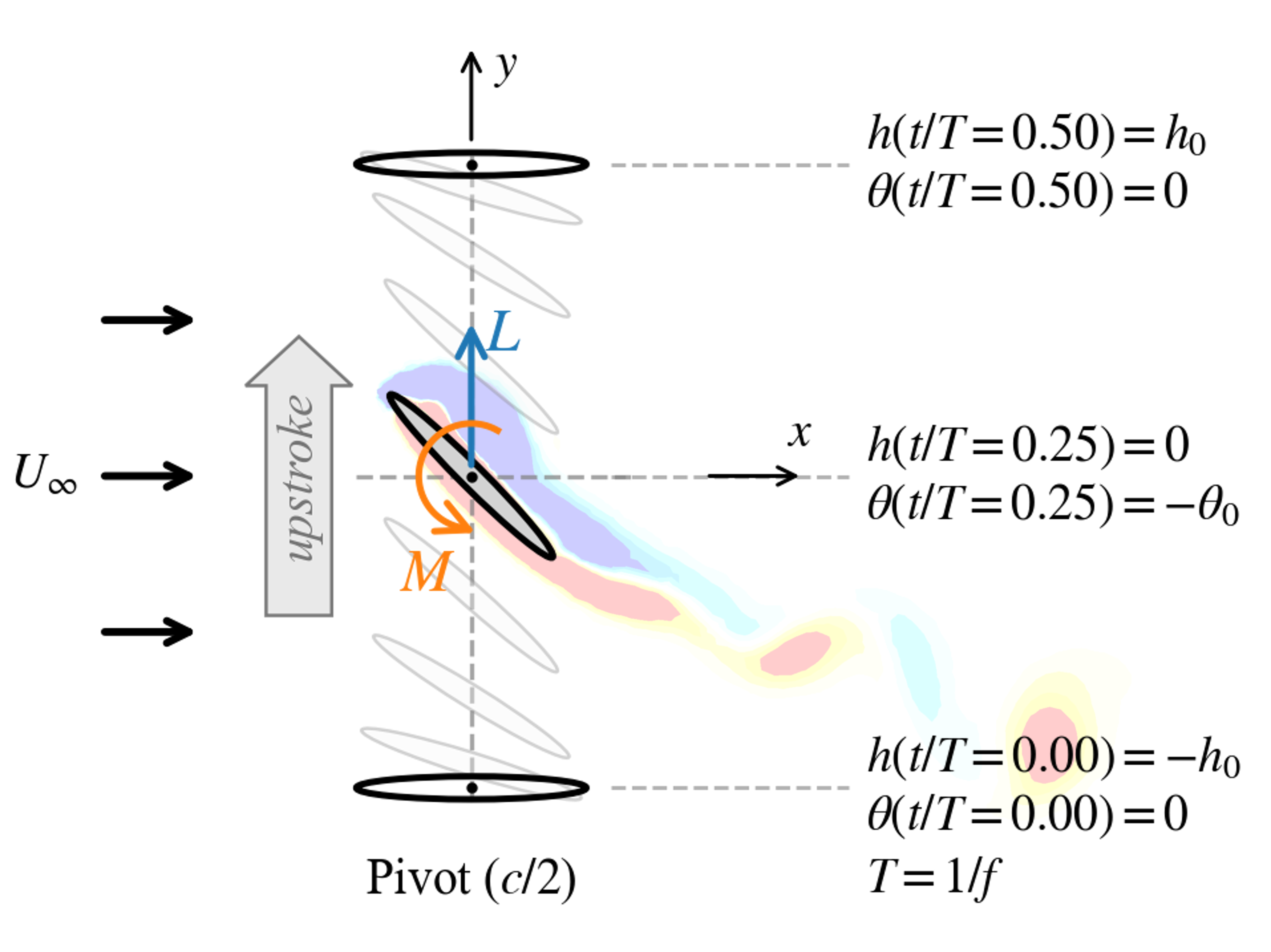}
    \end{subfigure}%
    \begin{subfigure}[t]{0.40\textwidth}
      \caption{}\label{fig:1b}
      \centering
      \includegraphics[width=1.00\linewidth]{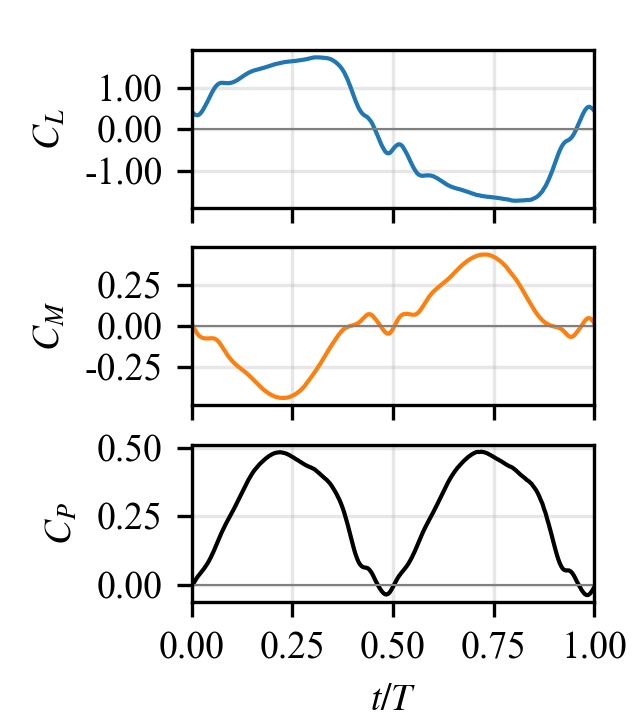}
    \end{subfigure}
    \caption{(\textit{a}) Schematic of the pitching--heaving foil over the upstroke half-cycle. The foil pivots about its mid-chord ($c/2$) in a uniform freestream $U_\infty$, generating lift $L$ and moment $M$. (\textit{b}) Instantaneous $C_L$, $C_M$, and $C_P$ over one full cycle at $(f, h_0, \theta_0) = (0.09, 0.50, 45^\circ)$.}
    \label{fig:sketch}
  \end{figure}

A two-dimensional foil with a 10\% thick elliptical cross-section undergoes prescribed sinusoidal heaving and pitching about its mid-chord ($c/2$), as depicted in figure~\ref{fig:1a}. The heave displacement $h(t)$ and pitch angle $\theta(t)$ are defined as
\begin{subequations}\label{eq:kinematics}
\begin{equation} \label{eq:heave}
    h(t) = -h_0 \cos(2\pi f t)
\end{equation}

and

\begin{equation} \label{eq:pitch}
  \theta(t) = -\theta_0 \sin(2\pi f t),
\end{equation}
\end{subequations}
where $h_0$ is the heave amplitude, $\theta_0$ is the pitch amplitude, and $f$ is the oscillation frequency, and all quantities are normalized by $c$ and $U_\infty$. Differentiating the prescribed motions in equation~\eqref{eq:kinematics} gives the non-dimensional heave and pitch velocity,
\begin{subequations}\label{eq:rates}
  \begin{equation}\label{eq:hdot}
    \dot h(t) = 2\pi f h_0 \sin(2\pi f t)
\end{equation}

and
  
  \begin{equation}\label{eq:thetadot}
    \dot\theta(t) = -2\pi f \theta_0 \cos(2\pi f t).
  \end{equation}
\end{subequations}

The unsteady response of the foil is characterized by the lift, pitching moment, and power coefficients
\begin{subequations}\label{eq:coeffs}
\begin{equation}\label{eq:lift_coeffs}
    C_L(t) = \frac{L(t)}{\tfrac12 \rho U_\infty^2 c},
\end{equation}
\begin{equation}\label{eq:moment_coeffs}
    C_M(t) = \frac{M(t)}{\tfrac12 \rho U_\infty^2 c^2},
\end{equation}
\end{subequations}
and 
\begin{equation}\label{eq:power}
  C_P(t) = C_{L}(t)\,\dot{h}(t) + C_{M}(t)\,\dot{\theta}(t),
\end{equation}
where $\rho$ is the fluid density, $L(t)$ is the lift force per unit span on the foil, and $M(t)$ is the pitching moment per unit span about the mid-chord pivot (positive counter-clockwise). The instantaneous power generated through the prescribed kinematics is $C_P(t)$, whereas its average over one oscillation cycle is denoted $\bar{C}_P$ and computed as
\begin{equation}\label{eq:cp_integral}
    \bar{C}_P = \frac{1}{T}\int_0^T
    \left( C_L\,\dot h + C_M\,\dot\theta \right) \mathrm{d}t .
\end{equation}
The energy-harvesting efficiency compares the extracted power with the kinetic-energy flux through the window swept by the foil,
\begin{equation}\label{eq:eta}
   \eta = \frac{\bar{P}}{\frac{1}{2}\rho U_\infty^3 Y_p} = \bar{C}_P\frac{c}{Y_p},
\end{equation}
where $\bar P$ is the cycle-averaged extracted power and $Y_p$ is the peak-to-peak transverse extent of the foil motion over one cycle, evaluated from the tip trajectory of the foil.

The kinematic setup and representative coefficients are illustrated in figure~\ref{fig:sketch}. The pitch extreme coincides with the mid-stroke of the heave motion, a $90^\circ$ phase offset that is the standard configuration for energy-harvesting foils \citep{KinseyDumas2008, RibeiroFranck2020}.
The effective angle of attack at the quarter-period, $\alpha_{T/4}$, is defined as~\citep{KinseyDumas2008}
\begin{equation}\label{eq:alpha_t4}
  \alpha_{T/4} = \theta_0 - \arctan(2\pi f h_0).
\end{equation}

To cover the energy-harvesting kinematic space, the parameters are sampled over $f \in \{0.08, 0.09, \ldots, 0.16\}$ (nine values), $h_0 \in \{0.50, 0.75, 1.00, 1.25, 1.50\}$ (five values), and $\theta_0 \in \{40^\circ, 45^\circ, \ldots, 80^\circ\}$ (nine values), giving $9 \times 5 \times 9 = 405$ kinematic combinations in total. This range covers the three flow regimes identified by \citet{Ribeiroetal2021}. This includes a shear-layer regime ($0^\circ < \alpha_{T/4} < 11.5^\circ$) where the flow remains attached and sheds only thin shear layers into the wake, an LEV regime ($11.5^\circ < \alpha_{T/4} < 28^\circ$) where the leading edge separates and rolls up into a strong LEV at each half-stroke, and an LEV+TEV regime ($\alpha_{T/4} > 28^\circ$) where the stronger incidence sheds an LEV and TEV pair. A subset of simulations at low and negative $\alpha_{T/4}$ produces net thrust instead of energy ($\bar{C}_P<0$) and is thus labeled as a propulsion regime. This propulsive data is included in the analysis but is not the focus of the current study. 

\begin{figure}
    \centering
    \begin{subfigure}[t]{0.5\textwidth}
      \caption{}\label{fig:2a}
      \centering
      \includegraphics[width=1.00\linewidth]{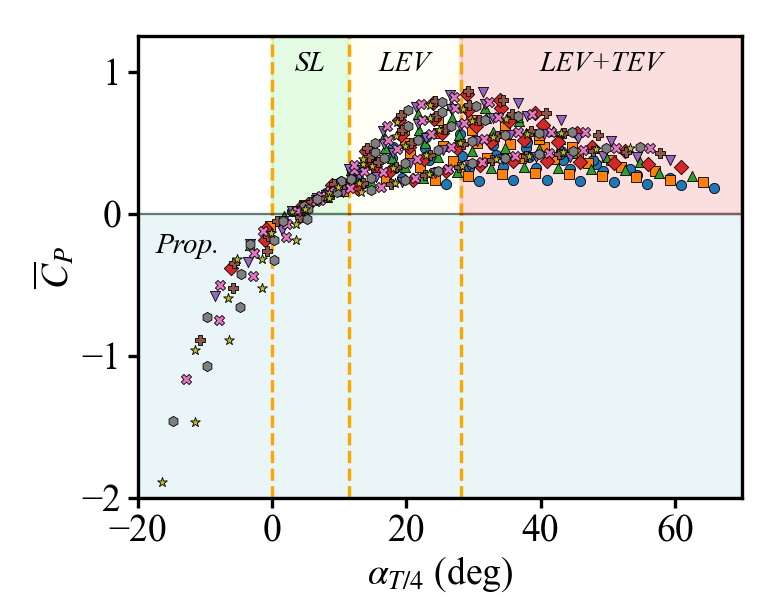}
    \end{subfigure}\hfill
    \begin{subfigure}[t]{0.5\textwidth}
      \caption{}\label{fig:2b}
      \centering
      \includegraphics[width=1.00\linewidth]{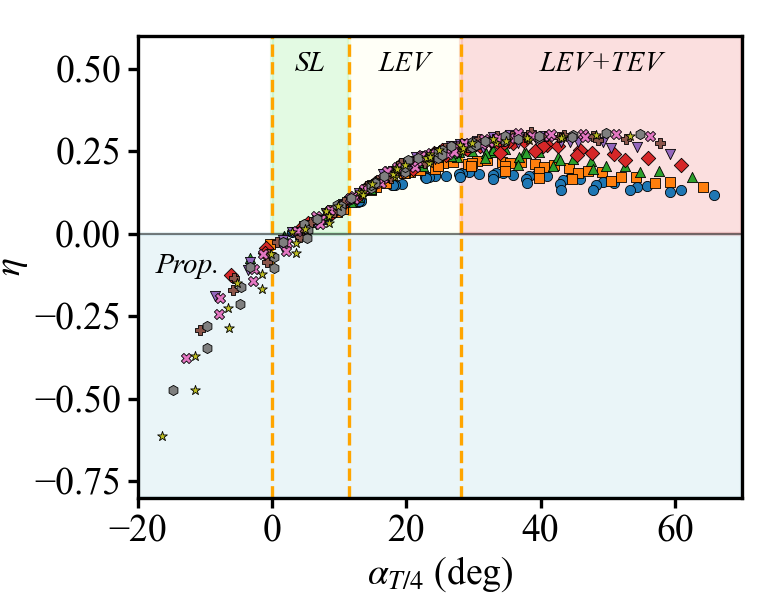}
    \end{subfigure}\\[3pt]
    \includegraphics[width=0.75\textwidth]{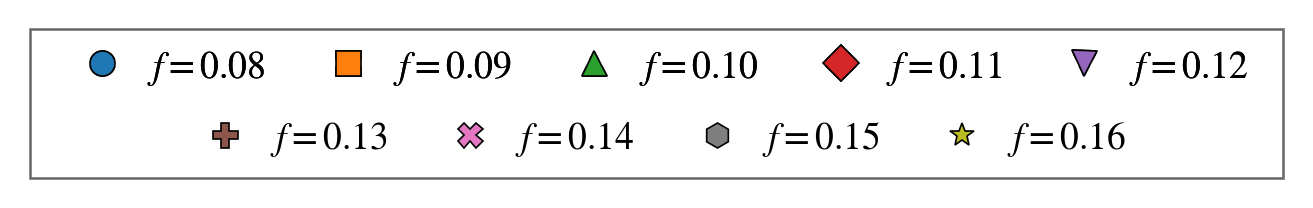}
    \caption{(\textit{a}) Cycle-averaged power coefficient $\bar{C}_P$ and (\textit{b}) energy-harvesting efficiency $\eta$
  versus $\alpha_{T/4}$. Shaded regions denote propulsion (blue), shear layer (green), LEV (yellow), and LEV+TEV (red).}
    \label{fig:distribution}
\end{figure}

The shaded bands in figure~\ref{fig:distribution} follow these boundaries, and show the power coefficient and efficiency associated with each regime. As seen in figure~\ref{fig:distribution}, the choice of $\alpha_{T/4}$ as a unifying kinematic parameter partially collapses the data.  The efficiency is approximately linear at small $\alpha_{T/4}$, but beyond roughly $20^\circ$ kinematics sharing the same $\alpha_{T/4}$ spread, primarily with respect to frequency. For mean power, $\bar{C}_P$, the spread begins earlier, around $\alpha_{T/4}\approx 10^{\circ}$.

\begin{figure}
    \centering
    \begin{subfigure}[t]{0.33\textwidth}
      \caption{}\label{fig:3a}
      \centering
      \includegraphics[width=1.00\linewidth]{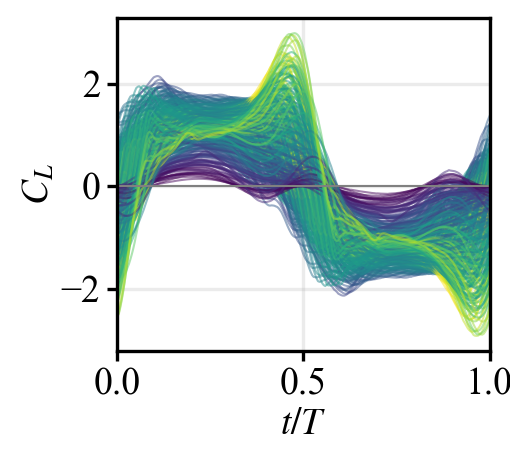}
    \end{subfigure}\hfill
    \begin{subfigure}[t]{0.33\textwidth}
      \caption{}\label{fig:3b}
      \centering
      \includegraphics[width=1.00\linewidth]{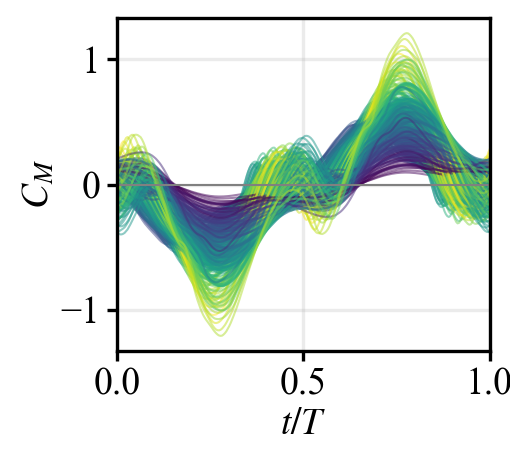}
    \end{subfigure}\hfill
    \begin{subfigure}[t]{0.33\textwidth}
      \caption{}\label{fig:3c}
      \centering
      \includegraphics[width=1.00\linewidth]{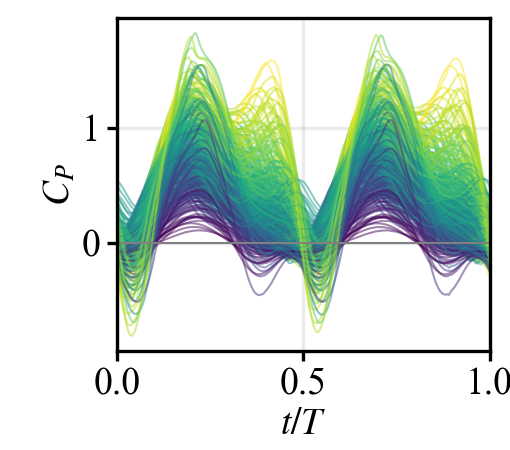}
    \end{subfigure}\\[0pt]
    \includegraphics[width=0.5\textwidth]{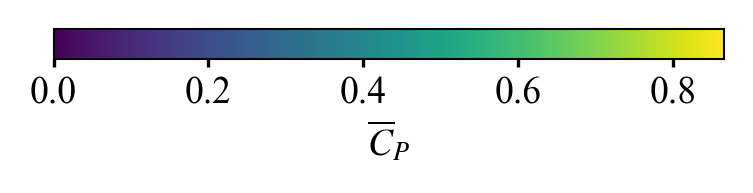}
    \caption{Instantaneous 
    (\textit{a}) lift, 
    (\textit{b}) pitching moment, and 
    (\textit{c}) power coefficients 
    over one cycle for all energy-harvesting simulations ($\bar{C}_P > 0$). Curves are colored by the mean power $\bar{C}_P$ of each simulation.}
    \label{fig:waveforms}
\end{figure}

The instantaneous power demonstrates the complexity of the relationship between kinematics and energy-harvesting performance. Figure~\ref{fig:waveforms} displays $C_{L}(t)$, $C_{M}(t)$, and $C_P(t)$ across all the simulations in the energy-harvesting regime ($\bar{C}_P>0$). As the produced power is composed from the lift and moment through the prescribed rates in equation~\eqref{eq:power}, any structure in $C_P$ originates in $C_L$ and $C_M$.  The lift in figure~\ref{fig:3a} ranges from near-sinusoidal curves at low amplitude to multi-peaked shapes reaching $|C_L|\approx3$, and the strongest harvesters concentrate their lift late in the stroke, just before the reversal at $t/T=0.5$ and 1. The moment coefficient in figure~\ref{fig:3b} is roughly a third of the lift in magnitude, reaches its extremes near mid-stroke, and carries visible higher-harmonic oscillations around the stroke reversals. Their products with the velocities generate a power curve that repeats every half cycle (figure~\ref{fig:3c}), with extraction concentrated at $t/T=0.25$ and 0.75 where the heave velocity is largest. The simulations of highest $\bar{C}_P$ also show a secondary peak just before the reversal, inherited from a late lift peak in figure~\ref{fig:3a}. 

One of the major conclusions from analyzing this data is that simulations with comparable $\alpha_{T/4}$ produce markedly different instantaneous responses, further motivating our methodology to discover better predictors of the time-dependent force and power curves. Another major observation is that the lift and moment inherit the periodicity of the input kinematics, which can be exploited by the Fourier decomposition framework for instantaneous prediction developed in \S\ref{sec:fourier}.
%



\section{Modeling the harmonic structure of power}
\label{sec:modeling}

\subsection{Fourier decomposition}
\label{sec:fourier}

\begin{figure}
    \centering
    \begin{subfigure}[t]{0.5\textwidth}
      \caption{}\label{fig:4a}
      \centering
      \includegraphics[width=1.00\linewidth]{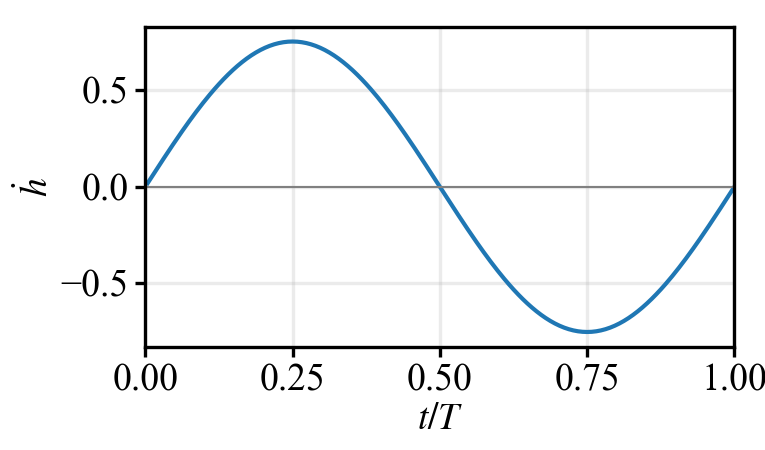}
    \end{subfigure}\hfill
    \begin{subfigure}[t]{0.5\textwidth}
      \caption{}\label{fig:4b}
      \centering
      \includegraphics[width=1.00\linewidth]{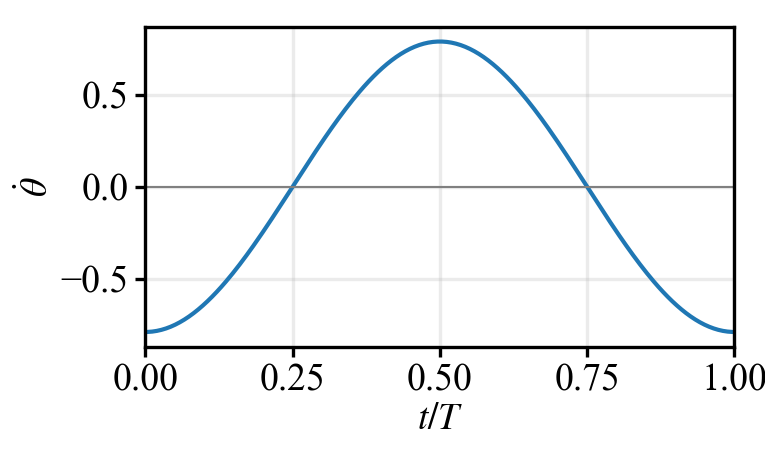}
    \end{subfigure}\\[3pt]
    \begin{subfigure}[t]{0.5\textwidth}
      \caption{}\label{fig:4c}
      \centering
      \includegraphics[width=1.00\linewidth]{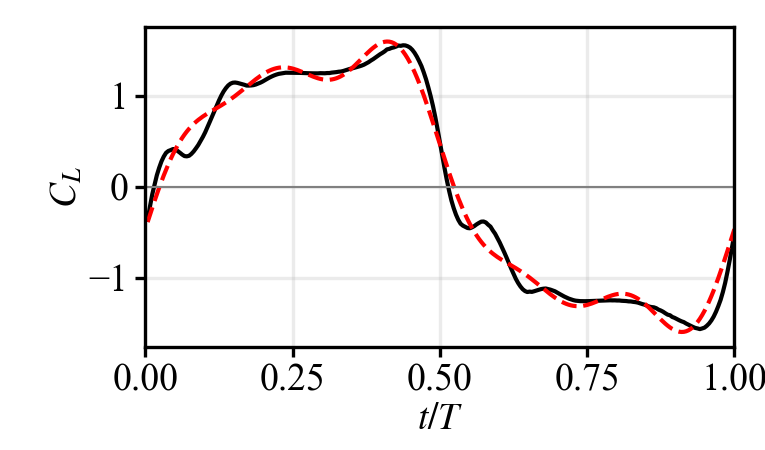}
    \end{subfigure}\hfill
    \begin{subfigure}[t]{0.5\textwidth}
      \caption{}\label{fig:4d}
      \centering
      \includegraphics[width=1.00\linewidth]{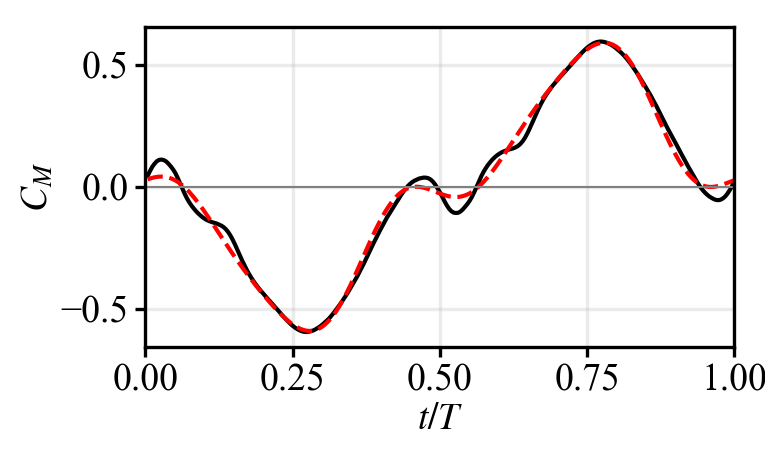}
    \end{subfigure}\\[3pt]
    \begin{subfigure}[t]{0.50\textwidth}
      \caption{}\label{fig:4e}
      \centering
      \includegraphics[width=1.00\linewidth]{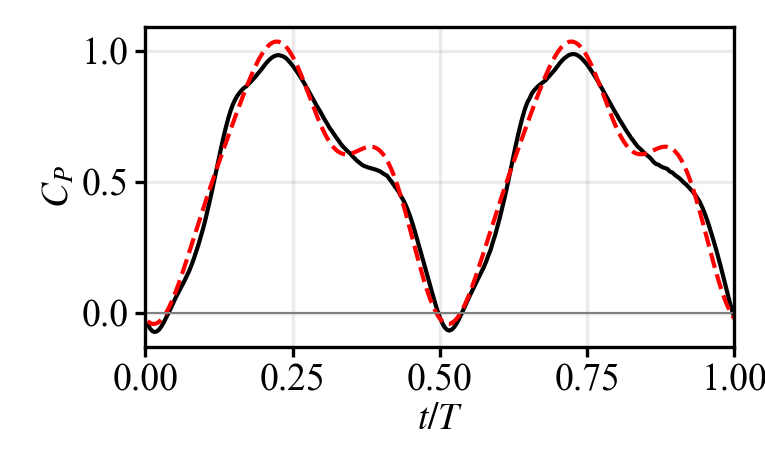}
    \end{subfigure}\\[3pt]
    \caption{Representative kinematic combination at $(f, h_0, \theta_0) = (0.12, 1.00, 60^\circ)$, showing
    (\textit{a}) the prescribed heave velocity,
    (\textit{b}) the prescribed pitch velocity,
    (\textit{c}) the lift coefficient,
    (\textit{d}) the pitching moment coefficient, and
    (\textit{e}) the power coefficient.
    In (\textit{c})--(\textit{e}), the simulated response (black solid) is overlaid with its $k=1,3,5$ Fourier reconstruction (red dashed).} 
    \label{fig:representative}
\end{figure}

\begin{figure}
    \centering
    \begin{subfigure}[t]{0.50\textwidth}
      \caption{}\label{fig:5a}
      \centering
      \includegraphics[width=1.00\linewidth]{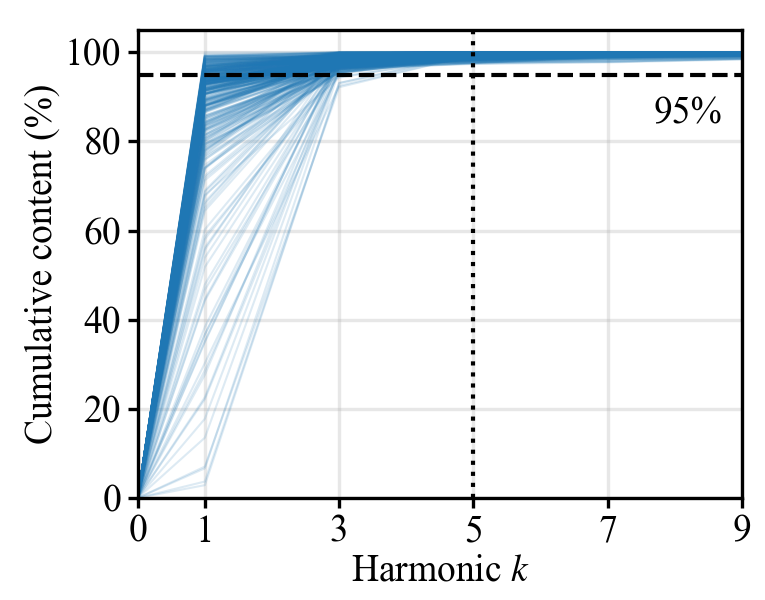}
    \end{subfigure}\hfill
    \begin{subfigure}[t]{0.50\textwidth}
      \caption{}\label{fig:5b}
      \centering
      \includegraphics[width=1.00\linewidth]{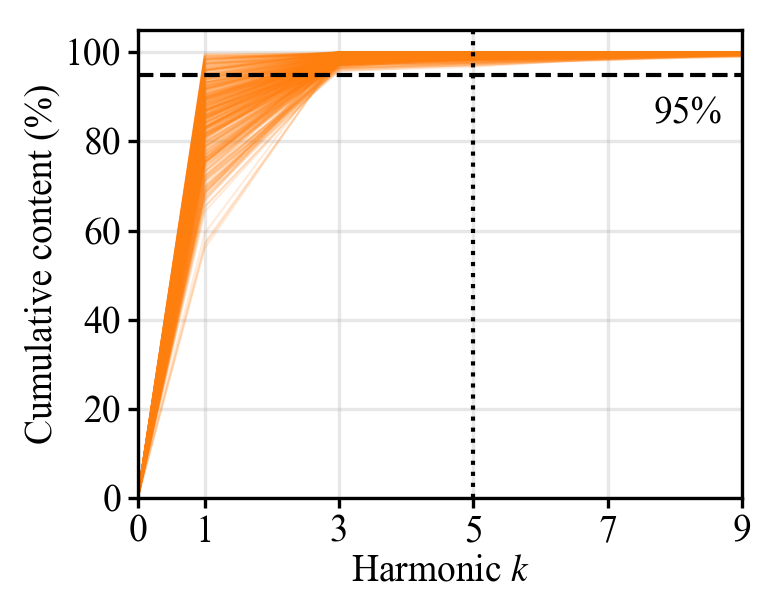}
    \end{subfigure}
    \caption{Cumulative spectral content of 
    (\textit{a}) $C_L$ and 
    (\textit{b}) $C_M$ 
    for all 405 simulations, shown at the odd harmonics ($k = 1, 3, 5, \ldots$), since even harmonics vanish. The truncation at $k = 5$ retains at least 95\% (dashed line) in every simulation.}
    \label{fig:energy_spectrum}
\end{figure}

A representative example of the prescribed velocities and the resulting lift force, pitch moment, and power coefficients is shown in figure~\ref{fig:representative}. 
%
Both the lift and the moment are periodic with the oscillation period $T$, so they can be represented as a Fourier series. The prescribed kinematics impose a half-cycle anti-symmetry, $h(t + T/2) = -h(t)$ and $\theta(t + T/2) = -\theta(t)$. Because the elliptical foil is symmetric about its chord line, the force and moment inherit this anti-symmetry, so that $C_{L}(t + T/2) =  -C_{L}(t)$ and $C_{M}(t + T/2) = -C_{M}(t)$. This eliminates all even harmonics ($k = 0, 2, 4, \ldots$), leaving only odd harmonics $k = 1, 3, 5, \ldots$. The truncation at $k = 5$ captures at least 95\% of the spectral content across all 405 simulations (figure~\ref{fig:energy_spectrum}), and the representations reduce to
\begin{subequations}\label{eq:fourier}
  \begin{equation} \label{eq:lift_fourier}
      C_L(t) \approx \sum_{k=1,3,5} \left[
      a_k^L \cos(2\pi k t / T) +
      b_k^L \sin(2\pi k t / T) \right],
  \end{equation}
and
  \begin{equation} \label{eq:moment_fourier}
      C_M(t) \approx \sum_{k=1,3,5} \left[
      a_k^M \cos(2\pi k t / T) +
      b_k^M \sin(2\pi k t / T) \right],
  \end{equation}
\end{subequations}
where the superscripts $L$ and $M$ distinguish the lift and moment coefficients, and $(a_k, b_k)$ are the cosine and sine coefficients of harmonic $k$. The strength of each harmonic is given by its amplitude $A_k = \sqrt{a_k^2 + b_k^2}$. Reconstructions from these three harmonics closely track the simulations in figures~\ref{fig:4c}--\ref{fig:4e}. Across all 405 simulations, the truncated reconstruction departs from the simulation by a root-mean-square difference of $0.10$ for $C_L$ and $0.026$ for $C_M$ at the median (at most $0.25$ and $0.061$), and shifts the peak lift and moment by no more than $12\%$, with a median value of $2\%$. Thus, the prediction model described below targets the six coefficients $(a_k, b_k)$ for $k = 1, 3, 5$ from both $C_{L}$ and $C_{M}$, yielding twelve prediction targets in total.
%


\subsection{Mean power from Fourier harmonics}
\label{sec:mean_power}

The mean power can easily be represented by the Fourier decomposition. 
The Fourier series in equation~\eqref{eq:fourier} for $C_L$ and $C_M$, and the heave and pitch velocities from equation~\eqref{eq:rates}, are substituted into equation~\eqref{eq:cp_integral}. Every term in the integrand is a product of two harmonics, and these are orthogonal over the period,
\begin{equation}\label{eq:ortho}
    \frac{1}{T}\!\int_0^T \!\sin(2\pi k t/T)\sin(2\pi t/T)\,\mathrm{d}t
    = \frac{1}{T}\!\int_0^T \!\cos(2\pi k t/T)\cos(2\pi t/T)\,\mathrm{d}t
    = \tfrac{1}{2}\,\delta_{k1},
\end{equation}
where $\delta_{k1}$ is the Kronecker delta, unity for $k = 1$ and zero otherwise. Every sine--cosine product also averages to zero, so only the $k = 1$ sine component of $C_L$ (paired with $\dot h$) and the $k = 1$ cosine component of $C_M$ (paired with $\dot\theta$) survive. The two terms from the heave and pitch motion then reduce to,
\begin{subequations}\label{eq:cp_channel}
    \begin{equation}\label{eq:cp_heave_channel}
        \frac{1}{T}\!\int_0^T C_L\,\dot h\,\mathrm{d}t = \pi f h_0\,b_1^{L},    
    \end{equation}
    and
    \begin{equation}\label{eq:cp_pitch_channel}
        \frac{1}{T}\!\int_0^T C_M\,\dot\theta\,\mathrm{d}t = -\pi f \theta_0\,a_1^{M},
    \end{equation}
\end{subequations}
and their sum gives the mean power,
\begin{equation}\label{eq:cp_bar} 
    \bar{C}_P = \pi f\left( b^{L}_1 h_0 - a^{M}_1 \theta_0 \right).
\end{equation}
In summary, only two Fourier coefficients $(b^{L}_1, a^{M}_1)$ are needed to predict $\bar{C}_P$. The other odd-harmonic coefficients of $C_L$ and $C_M$ shape the instantaneous response but average out over the cycle.

\subsection{Kinematic dependence of the Fourier harmonics}
\label{sec:harmonic}

\begin{figure}
    \centering
    \includegraphics[width=\textwidth]{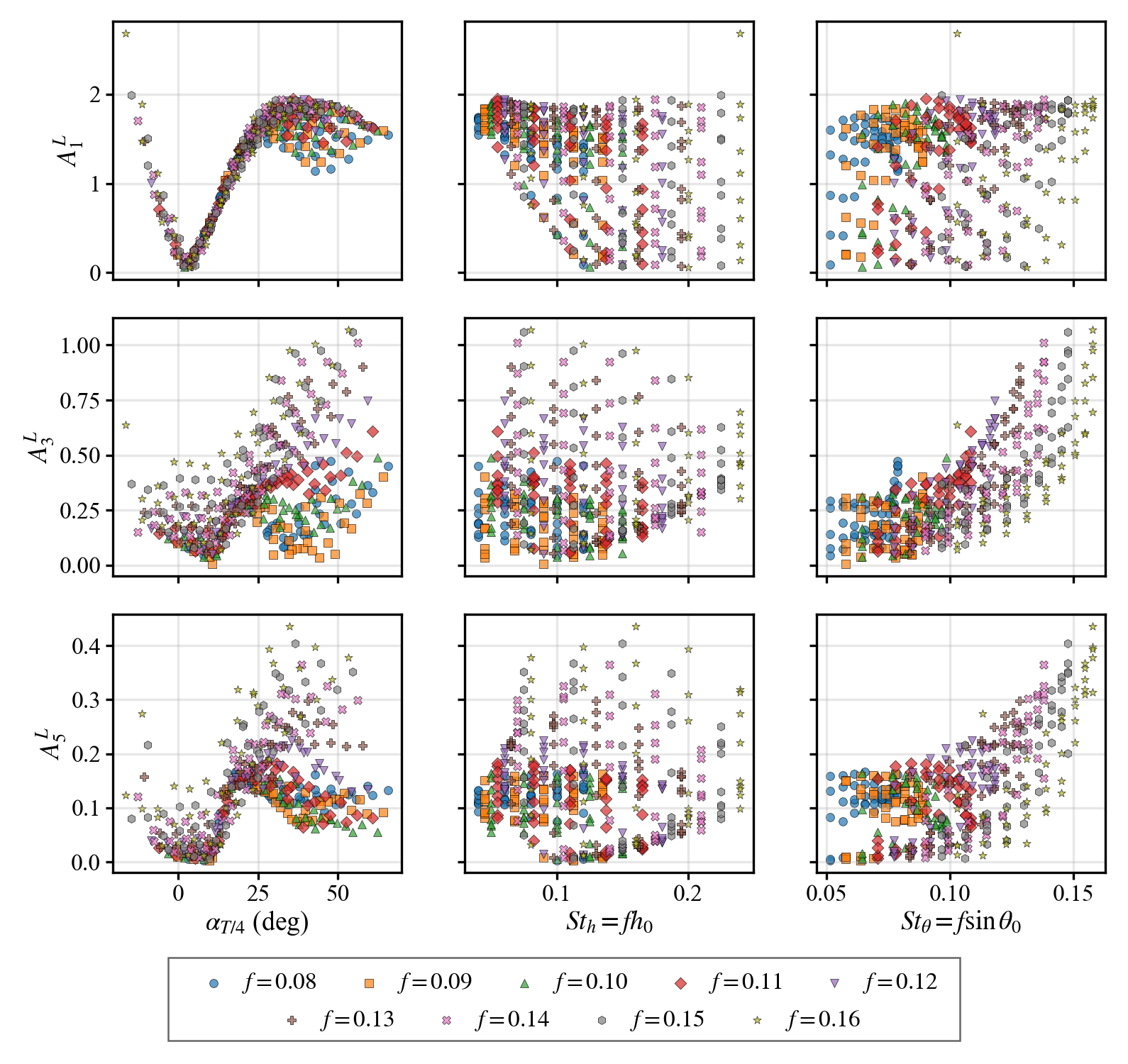}
    \caption{Harmonic amplitudes $A_k^L$ of the lift coefficient against three kinematic groups. Rows are $k = 1, 3, 5$ from top to bottom. Columns are the effective angle of attack $\alpha_{T/4}$, the heave Strouhal number $St_h = f h_0$, and the pitch Strouhal number $St_\theta = f \sin\theta_0$.}
    \label{fig:amplitude_scatter}
\end{figure}

In attempts to collapse the instantaneous force to established kinematics groups, the kinematic dependence of each Fourier harmonic of $C_L$ is examined. In the energy-harvesting literature the quarter-period angle $\alpha_{T/4}$ is widely used \citep{KinseyDumas2008, Kimetal2017}, whereas for propulsive regimes the wake structure and performance are often organized by a Strouhal number, the product of non-dimensional frequency and non-dimensional amplitude \citep{Triantafyllouetal1991, Andersonetal1998, Floryanetal2017}. For a foil that both heaves and pitches, there are at least two Strouhal numbers \citep{VanBurenetal2019}, $St_h = fh_0$ for heave and $St_\theta = f\sin\theta_0$ for the trailing-edge sweep of the pitch. To demonstrate how the harmonics of the current data scale with these traditional metrics, figure~\ref{fig:amplitude_scatter} plots the odd-harmonic amplitudes $A_k$ of $C_L$ ($k=1,3,5$) against $\alpha_{T/4}$, $St_h$, and $St_\theta$.

The amplitude of the first Fourier harmonic, $A^L_1$, collapses at small $\alpha_{T/4}$ up to $\alpha_{T/4}\approx 20^\circ$ before a bifurcation at high $\alpha_{T/4}$. This is  expected since the first harmonic largely captures the mean power, and similar behavior is seen for efficiency in figure~\ref{fig:2b}. In figure~\ref{fig:2a} the mean power spreads earlier, from about $10^\circ$, because it weights the first harmonic with the heave velocity $fh_0$ (\S\ref{sec:mean_power}), which itself varies at fixed $\alpha_{T/4}$. Neither Strouhal group captures trends with respect to $A^L_1$, as both scatter the points across the whole range. The higher harmonics have the opposite trend. For $\alpha_{T/4}$, $A^L_3$ and $A^L_5$ spread widely with only weak low-frequency clustering, yet there is a mild trend with $St_\theta$, in that the higher $St_{\theta}$ values correlate with higher $A^L_5$ amplitudes. The heave-based Strouhal number, $St_h$,  does not produce any noticeable trends with respect to the harmonic amplitude. The dominant grouping parameter therefore shifts with harmonic order, from $\alpha_{T/4}$ for the first harmonic to $St_\theta$ for the higher harmonics. 
%
%


\subsection{Scaling parameters and physics-derived features}
\label{sec:feature}

The previous section explored non-dimensional scalings from prior investigations, but none  are sufficient to capture the unsteady force and moment coefficients of the energy-harvesting regime. Thus, this section derives physics-based terms that govern kinematics, dynamics, and force production. 
These terms will be utilized as features in the Ridge regression model introduced in the following section. 

The kinematic stroke is governed by $(f, h_0, \theta_0)$, which provide three independent parameters, and the lowest-order features. The heave and pitch velocities in equation~\eqref{eq:rates} scale as $|\dot h|\propto fh_0$ and $|\dot\theta|\propto f\theta_0$. The acceleration terms are then
\begin{subequations}
\begin{equation}
    \ddot h(t) = (2\pi f)^2 h_0 \cos(2\pi f t),
\end{equation}
and
\begin{equation}
  \ddot\theta(t) = (2\pi f)^2 \theta_0 \sin(2\pi f t),
\end{equation}
\end{subequations}
providing $|\ddot h|\propto f^2 h_0$, $|\ddot\theta|\propto f^2\theta_0$, both of which have $f^2$ dependence. As will be confirmed by the results, $f^2$ is influential independently of $h_0$ and $\theta_0$, and can be derived from the spatial gradient of the acceleration, $d\ddot{h}/dh$ and $d\ddot{\theta}/d\theta$, both of which scale with $f^2$ at leading order.

The quasi-steady lift can be approximated through thin-airfoil theory as $C_{L}\approx 2\pi\alpha_\mathrm{eff}$, where
\begin{equation}
  \alpha_\mathrm{eff}(t) = -\theta(t) - \arctan(\dot h(t)).
\end{equation}
Substituting equations~\eqref{eq:pitch} and \eqref{eq:hdot}, and using a small angle approximation for $\arctan$ reveals leading order terms of $\theta_0$ and $fh_0$, which are already accounted for in the kinematics scaling, and neglects the third order term $(fh_0)^3$.
%
%
The quasi-steady moment from thin-airfoil theory, $C_{M}\approx -\tfrac14 C_{L}$, is also accounted for since it inherits the same scaling. 

The added-mass terms contributing to the lift \citep{Theodorsen1935} are
\begin{equation}
  C_{L}^\mathrm{am} = \tfrac12\pi\,[\ddot h + \dot\theta]
  \end{equation}
whereas for the moment about $c/2$ they are 
\begin{equation}
  C_{M}^\mathrm{am} = -\tfrac{1}{8}\pi\,[\dot\theta + \tfrac18\ddot\theta],
\end{equation}
repeating terms already derived from the velocity and acceleration scaling. 

Starting with equation~\eqref{eq:power} and substituting the velocities from equation~\eqref{eq:rates} provides the amplitude scaling for heave power as $|C_{P_h}|\propto(fh_0\theta_0,\, f^2 h_0^2,\, f^2 h_0\theta_0)$. A similar process follows for pitch power, which reveals $|C_{P_\theta}|\propto(f\theta_0^2,\, f^2 h_0\theta_0,\, f^2\theta_0^2)$. Together these yield five features distinct to power.

The energy extracted per cycle is computed as force times displacement, or $E_h = C_{L} h_0$. Multiplying the heave component by those already derived for $C_L$ yields $|E_h|\propto (fh_0^2,\ h_0\theta_0,\ f^2h_0^2,\ fh_0\theta_0)$. The rotational energy per cycle is derived similarly, and scales with $|E_\theta|\propto (\theta_0^2,\ f\theta_0^2,\ f^2\theta_0^2,\ fh_0\theta_0)$.

Table~\ref{tab:features} summarizes all sixteen terms, their polynomial order, and their physical origin from the kinematics and dynamic scaling analysis. The terms are arranged into six groups by their variable content (linear, quadratic, frequency--heave, frequency--pitch, heave--pitch, and three-parameter), and the same grouping organizes the attribution analysis of \S\ref{sec:step1_ridge}.
\begin{table}
\begin{center}
    \renewcommand{\arraystretch}{1.3}
    \setlength{\tabcolsep}{4pt}
    \resizebox{\textwidth}{!}{%
    \begin{tabular}{ccccl}
        \shortstack{Feature \\ group} & \shortstack{Feature \\ variable $x_j$} & Term
              & \shortstack{Polynomial \\ order} & Physical origin \\
        \hline
        Linear
          & $x_1$    & $f$               & 1 & Kinematic input \\
          & $x_2$    & $h_0$             & 1 & Kinematic input \\
          & $x_3$    & $\theta_0$        & 1 & Kinematic input + quasi-steady force and moment \\
        \hline
        Quadratic
          & $x_4$    & $f^2$             & 2 & Spatial gradient of heave/pitch acceleration \\
          & $x_5$    & $\theta_0^2$      & 2 & Energy from quasi-steady moment \\
        \hline
        Frequency--heave
          & $x_6$    & $fh_0$            & 2 & Heave velocity + quasi-steady force and moment \\
          & $x_7$    & $f^2 h_0$         & 3 & Heave acceleration + added-mass force \\
          & $x_8$    & $fh_0^2$          & 3 & Energy from quasi-steady force \\
          & $x_9$    & $f^2 h_0^2$       & 4 & Power from heave + energy from added-mass force \\
        \hline
        Frequency--pitch
          & $x_{10}$ & $f\theta_0$       & 2 & Pitch velocity + added-mass force and moment \\
          & $x_{11}$ & $f^2\theta_0$     & 3 & Pitch acceleration + added-mass moment \\
          & $x_{12}$ & $f\theta_0^2$     & 3 & Power from pitch + energy from added-mass moment \\
          & $x_{13}$ & $f^2\theta_0^2$   & 4 & Power from pitch + energy from added-mass moment \\
        \hline
        Heave--pitch
          & $x_{14}$ & $h_0\theta_0$     & 2 & Energy from quasi-steady force \\
        \hline
        Three-parameter
          & $x_{15}$ & $fh_0\theta_0$    & 3 & Power from heave + energy from quasi-steady moment \\
          & $x_{16}$ & $f^2 h_0\theta_0$ & 4 & Power from heave and pitch + energy from added-mass force \\
    \end{tabular}}
    \caption{Complete set of 16 physics-derived terms used as input to the prediction model, grouped by variable content to match the blocks of figure~\ref{fig:ridge_heatmap}.}
    \label{tab:features}
    \end{center}
\end{table}

%

\subsection{Setup of prediction model}
\label{sec:ridge}

Section~\ref{sec:feature} derived a library of 16 terms from products and powers of $(f, h_0, \theta_0)$. Each of the twelve Fourier coefficients is regressed onto this library, whose columns are the terms $x_j$ ($j=1,\dots,16$) listed in table~\ref{tab:features} and are features of the regression. The prediction model is built from this library in two steps with two distinct goals. Step~1 fits each of the twelve Fourier coefficients in a basis where the features are only weakly correlated, so the weight magnitudes can be read as the contribution of each term, and thereby identifies the most important terms that form each Fourier coefficient. Step~2 refits only the retained terms in the original monomial basis, so that the model acts on $(f, h_0, \theta_0)$ directly, and delivers the closed-form analytical model.

\noindent{\textbf{\textit{Step 1:}}} 

In Step~1, since the physics-derived features are correlated the raw regression weights do not directly indicate the importance of each feature. To better quantify their attribution, the features are recast in a Legendre-polynomial basis. Each kinematic parameter $q \in \{f, h_0, \theta_0\}$ is linearly mapped to the interval $[-1, 1]$,
\begin{equation}\label{eq:mapping}
    \tilde{q} = \frac{2(q - q_{\min})}{q_{\max} - q_{\min}} - 1,
\end{equation}
where $q_{\min}$ and $q_{\max}$ are the bounds of that parameter in the sweep. The Legendre polynomials up to degree 2, applied to any mapped parameter $\tilde{q}$, are
\begin{equation}\label{eq:legendre_poly}
  P_0(\tilde{q}) = 1, \quad 
  P_1(\tilde{q}) = \tilde{q}, \quad 
  P_2(\tilde{q}) = \tfrac{1}{2}(3\tilde{q}^2 - 1).
\end{equation}
Each term $x_j = f^a h_0^b \theta_0^c$ is then replaced by its Legendre-transformed counterpart
\begin{equation}\label{eq:legendre_feature}
  \phi_j = P_a(\tilde{f})\,P_b(\tilde{h}_0)\,
           P_c(\tilde{\theta}_0).
\end{equation}
This transform reduces the largest pairwise correlation among the features from $0.996$ to $0.37$, so that the transformed features are much closer to orthogonal. The Legendre polynomials stay within $\pm1$, so the features are normalized and share a common scale. Step~1 fits the linear model
\begin{equation}\label{eq:ridge_normalized}
    \hat y_i = \beta_0 + \sum_{j=1}^{p} \beta_j \,\phi_{j,i},
    \qquad i = 1,\ldots,N,
\end{equation}
where $\phi_{j,i}$ is feature $j$ at sample $i$ and $\hat y_i$ the predicted value of $y$. The weights are fitted by Ridge regression \citep{HoerlKennard1970},
\begin{equation}\label{eq:ridge_step1}
    (\hat{\beta}_0,\, \hat{\boldsymbol{\beta}}) = \arg\min_{\beta_0,\, \boldsymbol{\beta}}\,
    \sum_{i=1}^{N} \Big( y_i - \beta_0 - \sum_{j=1}^{p} \beta_j\, \phi_{j,i} \Big)^{2}
    + \lambda \sum_{j=1}^{p} \beta_j^{2},
\end{equation}
where $y_i$ is the target value at sample $i$ and $\lambda \geq 0$ is the regularization strength, selected by leave-one-out cross-validation. The penalty excludes the intercept and shrinks the remaining weights, so the attribution does not rest on weakly determined ones. The magnitude $|\beta_j|$ is a first measure of how strongly feature $j$ drives the target (\S\ref{sec:step1_ridge}).

For a fair comparison across features, the attribution uses the standardized regression weight $\tilde{\beta}_j = \beta_j\,\sigma(\phi_j)/\sigma(y)$, where $\sigma(\cdot)$ is the sample standard deviation across the dataset of 405 simulations. It measures the change in the target, in units of its standard deviation, produced by a one-standard-deviation change in feature $j$. This correction matters because the product features have systematically smaller spread than the single-parameter features.

The standardized magnitudes $|\tilde{\beta}_j|$ are comparable within a target, but their absolute scales differ across the twelve targets ($a_k$ or $b_k$ of $C_{L}$ or $C_{M}$, $k \in \{1,3,5\}$). For cross-target comparison, each target's $|\tilde{\beta}_j|$ is normalized to sum to 100\%, 
\begin{equation}\label{eq:ridge_pct}    
    w_j = \frac{\left|\tilde{\beta}_j\right|}{\sum_{j'=1}^{p}\left| \tilde{\beta}_{j'}\right|} \times 100\%,
  \qquad \sum_{j=1}^{p} w_j = 100\%,
\end{equation}           
where the sums run over the $p$ features in equation~\eqref{eq:ridge_normalized} and exclude the intercept.

\noindent{\textbf{\textit{Step 2:}}} 

Step~2 turns the attribution into a directly deployable predictor. From the Step~1 contributions $w_j$, it retains for each target only the features whose cumulative contribution reaches $95\%$ of the total, and refits them in the original monomial basis, since the predictor must act on $(f, h_0, \theta_0)$ directly.

Unlike the Legendre features, the raw monomials span very different magnitudes, so a uniform Ridge penalty would unfairly shrink the small-scale ones. Weighting each penalty term by the feature's maximum $\max_i|x_{j,i}|$ removes this bias, and the weight vector $\mathbf{g}_m$ for target $m$ minimizes
\begin{equation}\label{eq:step2_gmj}
    \mathbf{g}_m = \arg\min_{\mathbf{g}} \left\| \mathbf{y} - \mathbf{X}\mathbf{g} \right\|_2^2
    + \lambda \sum_{j} \big( g_{j} \max_i|x_{j,i}| \big)^2,
\end{equation}
with $\mathbf{X}$ the raw feature matrix, $\mathbf{y}$ the vector of target values, $g_{m,j} = 0$ on the discarded features, and $\lambda$ again selected per target by leave-one-out cross-validation.

For the lift, the six weight vectors stack row by row into a matrix $\mathbf{G}^{L}$ with elements $g^{L}_{m,j}$, where $m \in \{a_1, b_1, a_3, b_3, a_5, b_5\}$ indexes the six Fourier coefficients and $j = 1, \ldots, 16$ the features, and the moment gives an analogous matrix $\mathbf{G}^{M}$ with elements $g^{M}_{m,j}$. The model has no intercept, so that zero kinematic input produces zero force and moment.

The predictor maps the feature vector $\mathbf{x} = (x_1, \ldots, x_{16})^{\top}$ to the Fourier coefficients through a single matrix--vector product,
\begin{subequations}\label{eq:closed_form}
    \begin{equation}\label{eq:closed_form_lift}
        \mathbf{c}^{L} = \mathbf{G}^{L}\mathbf{x},
    \end{equation}
    and
    \begin{equation}\label{eq:closed_form_moment}
        \mathbf{c}^{M} = \mathbf{G}^{M}\mathbf{x},
    \end{equation}
\end{subequations}
where $\mathbf{c}^{L} = (a_1, b_1, a_3, b_3, a_5, b_5)^{\top}$ collects the lift Fourier coefficients and $\mathbf{c}^{M}$ the moment Fourier coefficients. Substituting these coefficients into the Fourier series~\eqref{eq:fourier} recovers $C_{L}(t)$ and $C_{M}(t)$, from which $C_P(t)$ is computed (\S\ref{sec:expression}).

\section{Results}
\label{sec:results}

\subsection{Step~1 Ridge regression model}
\label{sec:step1_ridge}


The fit quality of Step~1 Ridge regression is measured by the leave-one-out coefficient of determination (LOO $R^2$), 
\begin{equation}
    R^2_\mathrm{LOO} \;=\; 1 - \frac{\sum_{i=1}^{N}\left(y_i - \hat y_i^{(-i)}\right)^2}{\sum_{i=1}^{N}\left(y_i - \bar y\right)^2},
    \label{eq:loo_r2}
\end{equation}
where $y_i$ denotes the reference target for the $i$th sample, $\hat y_i^{(-i)}$ is the prediction for sample $i$ from a surrogate trained on the remaining $N-1$ samples, $\bar{y}$ is the sample mean, and $R^2_\mathrm{LOO}=1$ indicates perfect generalization. 

Ten of the twelve target coefficients have an LOO $R^2 \geq 0.79$, and the two coefficients that contribute to the mean power ($b_1^{L}$ and $a_1^{M}$) are $0.99$ and $0.93$.  The moment coefficients all have an LOO $R^2 \geq 0.80$. The two weakest targets correspond to the $k=5$ Fourier coefficients of $C_{L}$ ($R^2 = 0.64$ for $a_5^{L}$ and $0.46$ for $b_5^{L}$). These coefficients are tied to the vortex shedding events which are examined in \S\ref{sec:shedding}. 



\begin{figure}
    \centering
    \begin{subfigure}{\textwidth}\caption{}\label{fig:7a}
      \includegraphics[width=\linewidth]{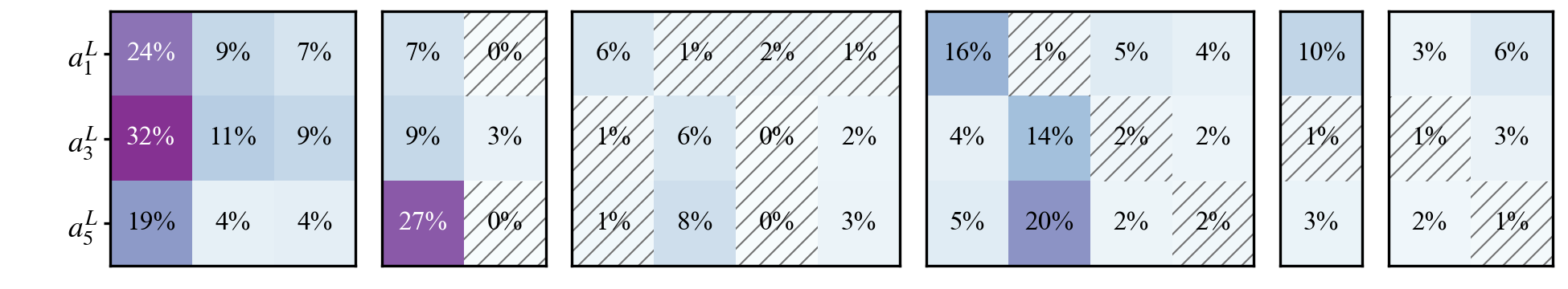}\end{subfigure}
    \begin{subfigure}{\textwidth}\caption{}\label{fig:7b}
      \includegraphics[width=\linewidth]{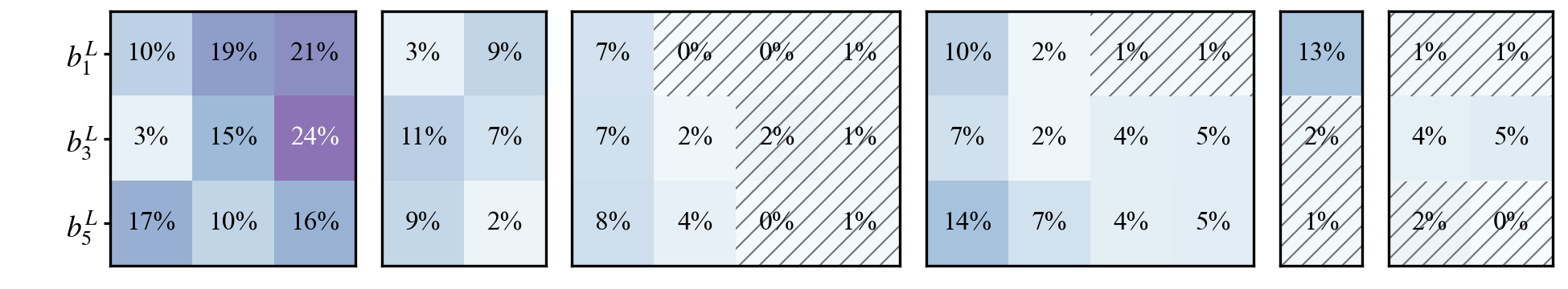}\end{subfigure}
    \begin{subfigure}{\textwidth}\caption{}\label{fig:7c}
      \includegraphics[width=\linewidth]{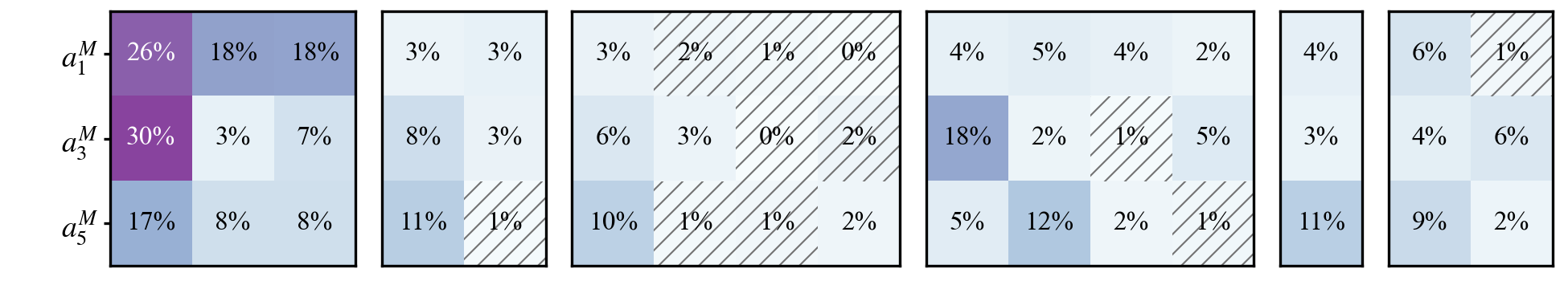}\end{subfigure}
    \begin{subfigure}{\textwidth}\caption{}\label{fig:7d}
      \includegraphics[width=\linewidth]{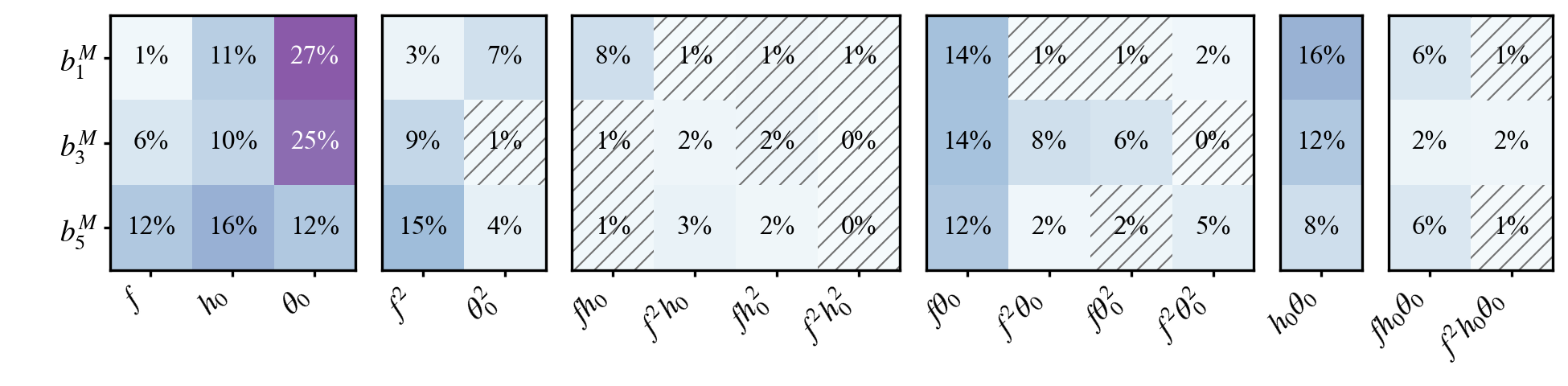}\end{subfigure}
    \includegraphics[width=0.5\textwidth]{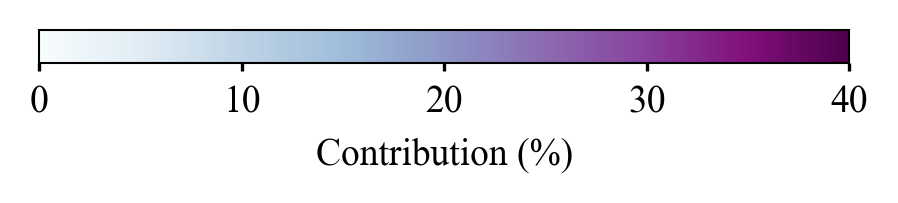}
    \caption{Step~1 Ridge attribution of (\textit{a}) $a_k^L$, (\textit{b}) $b_k^L$, (\textit{c}) $a_k^M$, (\textit{d}) $b_k^M$ ($k=1,3,5$). Each cell gives the normalized contribution $w_j$ (\%) of equation~\eqref{eq:ridge_pct}. Features are ordered and grouped as in table~\ref{tab:features}. Each feature is the Legendre product of equation~\eqref{eq:legendre_feature}, labeled by the monomial with the same exponents. Hatched cells mark the features discarded by the 95\% selection of Step~2, corresponding to the empty cells of table~\ref{tab:coef_step2}.
    }
    \label{fig:ridge_heatmap}
\end{figure}

The heatmaps in figure~\ref{fig:ridge_heatmap} show the percentage contribution of each feature to each Fourier coefficient, and reveal a systematic shift of the attribution with respect to harmonic order. Averaged across the four panels in figures~\ref{fig:7a}--\ref{fig:7d}, the linear set of features has $48\%$ of the contribution for $k=1$, but decreases slightly to $36\%$ at $k=5$. Likewise, the heave--pitch term decreases from $11\%$ to $6\%$ between $k=1$ and $k=5$. In contrast, the quadratic group contributes only $9\%$ at $k=1$, and increases to $17\%$ for $k=5$. Likewise, the frequency--pitch group increases from $18\%$ to $25\%$. Within these two groups the growth concentrates on the terms quadratic in frequency, as $f^2$ increases from $4\%$ to $16\%$ and $f^2\theta_0$ from $2\%$ to $10\%$. Section~\ref{sec:shedding} traces this growing dependence on frequency to the timing of the vortex shedding within each stroke.


The $a$ and $b$ coefficients are driven by features with different phase character, reflecting the natural sine-phase and cosine-phase split of the kinematic input. Sine-phase content (carried by $\theta_0$ and $fh_0$ in the quasi-steady response) is evident in the $b$ coefficients (figures~\ref{fig:7b} and~\ref{fig:7d}), which are led by $\theta_0$ ($21\%$ for $b_1^{L}$, $24\%$ for $b_3^{L}$). The frequency--pitch group has a rapid increase in the $b^L$ coefficients from $14\%$ at $k=1$ to $30\%$ at $k=5$. Cosine-phase content (mainly from added-mass terms) is prominent in $a$ coefficients, weighted toward frequency-bearing features, such as the $f^2$ term contributing $27\%$ for $a_5^{L}$ (figures~\ref{fig:7a} and~\ref{fig:7c}).

For the coefficients contributing to the mean power ($b_1^{L}$ in figure~\ref{fig:7b} and $a_1^{M}$ in figure~\ref{fig:7c}), the linear group contains $50\%$ and $63\%$ of the attribution, with the frequency-bearing groups at $27\%$ and $20\%$. This is consistent with the single $\alpha_{T/4}$ trend of the mean power at small angles (figure~\ref{fig:2a}). The heave--pitch term $h_0\theta_0$ contributes $10$--$13\%$ of the first-harmonic coefficients and under $3\%$ at $k=5$. This term is an energy cross term that feeds the mean power rather than the higher harmonics.

These attribution patterns determine which features are retained in the Step~2 closed-form predictor of \S\ref{sec:expression}. The retained set for each coefficient is shown in figure~\ref{fig:ridge_heatmap} and listed in table~\ref{tab:coef_step2}.
%
%
%


\subsection{Analytical expression from the Step~2 Ridge regression model}
\label{sec:expression}

The Step~2 predictor~\eqref{eq:closed_form} of \S\ref{sec:ridge} expresses each Fourier coefficient as a weighted sum of the retained features in $(f, h_0, \theta_0)$. For $C_L$, the prediction is a single matrix--vector product,
\begin{equation}\label{eq:closed_form_matrix}
      \begin{bmatrix} a_1^{L}\\[4pt] b_1^{L}\\[4pt] a_3^{L}\\[4pt] b_3^{L}\\[4pt] a_5^{L}\\[4pt] b_5^{L} \end{bmatrix}
      =
      \underbrace{\begin{bmatrix}
          g_{a_1, 1}^L & g_{a_1, 2}^L & \cdots & g_{a_1, 16}^L\\
          g_{b_1, 1}^L & g_{b_1, 2}^L & \cdots & g_{b_1, 16}^L\\
          \vdots     & \vdots     & \ddots & \vdots\\
          g_{b_5, 1}^L & g_{b_5, 2}^L & \cdots & g_{b_5, 16}^L\\
      \end{bmatrix}}_{\mathbf{G}^{L},\,6\times16}
      \underbrace{\begin{bmatrix}
          f\\ h_0\\ \theta_0\\ f^2\\ \theta_0^2\\ \vdots\\ fh_0\theta_0\\ f^2 h_0\theta_0
      \end{bmatrix}}_{16\times1}
\end{equation}
where $\mathbf{G}^{L}$ contains the Step~2 Ridge weights for $C_L$. Each row of $\mathbf{G}^{L}$ is nonzero only on the features Step~2 retained. The moment weights form an analogous matrix $\mathbf{G}^{M}$, and the two matrices ($\mathbf{G}^{L}$ and $\mathbf{G}^{M}$) together form the trained model, with the final coefficients listed in table~\ref{tab:coef_step2}. The same two steps produce a third matrix $\mathbf{G}^{D}$ for the streamwise force (appendix~\ref{sec:appendix_drag}). 

%
%
\begin{table}
\begin{center}
    \renewcommand{\arraystretch}{1.2}
    \setlength{\tabcolsep}{3.5pt}
    \resizebox{\textwidth}{!}{%
    \begin{tabular}{l|cccccc|cccccc}
        & \multicolumn{6}{|c|}{$\mathbf{G}^L$} & \multicolumn{6}{c}{$\mathbf{G}^M$} \\
        Feature & $a_1^L$ & $b_1^L$ & $a_3^L$ & $b_3^L$ & $a_5^L$ & $b_5^L$ & $a_1^M$ & $b_1^M$ & $a_3^M$ & $b_3^M$ & $a_5^M$ & $b_5^M$ \\
        \hline
        $f$ & -27.8 & 11.3 & -29.8 & 0.995 & -1.34 & -6.43 & -0.537 & -0.467 & 0.203 & 7.69 & -2.85 & -0.111 \\
        $h_0$ & 1.47 & -2.63 & -0.103 & 1.51 & 0.0582 & 0.762 & 1 & -1.08 & -0.176 & -0.29 & 0.318 & 0.139 \\
        $\theta_0$ & 2.29 & 4.48 & 1.51 & -4.67 & -1.19 & -4.13 & -2.32 & -0.278 & 0.265 & -0.789 & -0.0214 & -0.244 \\
        \hline
        $f^2$ & 140 & -171 & 216 & 48.1 & 16.8 & 57.4 & 6.35 & -3.45 & 4.29 & -51.1 & 27.6 & 0.507 \\
        $\theta_0^2$ & -- & -4.67 & -1.93 & 1.46 & -- & 2.61 & 1.99 & 0.543 & -0.0113 & -- & -- & 0.0679 \\
        \hline
        $f h_0$ & -4.03 & -16.8 & -- & -40.6 & -- & -12.1 & -8.27 & 20.3 & -2.67 & -- & -3.16 & -- \\
        $f^2 h_0$ & -- & -- & -3.75 & 219 & -1.47 & 42.1 & -- & -- & 28.1 & 0.0798 & -- & -3.83 \\
        $f h_0^2$ & -- & -- & -- & -- & -- & -- & -- & -- & -- & -- & -- & -0.156 \\
        $f^2 h_0^2$ & -- & -- & 2.88 & -- & -2.52 & -- & -- & -- & -- & -- & -0.728 & -- \\
        \hline
        $f \theta_0$ & -3.38 & 28.5 & 44.6 & 93 & 23.2 & 84.1 & 30.9 & -8.56 & -1.14 & 5.92 & 4.21 & 3.4 \\
        $f^2 \theta_0$ & -- & 32.7 & -392 & -551 & -118 & -408 & -90.3 & -- & -18.9 & -2.21 & -38.9 & -15.4 \\
        $f \theta_0^2$ & 13.4 & -- & -- & -36.4 & 0.324 & -52.7 & -29.2 & -- & -- & 0.114 & 0.0995 & -- \\
        $f^2 \theta_0^2$ & -138 & -- & 92.2 & 283 & -- & 258 & 89.3 & 20 & 4.68 & -- & -- & 1.42 \\
        \hline
        $h_0 \theta_0$ & -0.77 & 3.34 & -- & -- & -0.183 & -- & -0.925 & 0.673 & 0.43 & 0.486 & -0.445 & -0.234 \\
        \hline
        $f h_0 \theta_0$ & -21.1 & -- & -- & 14.8 & 1.77 & -- & 8.7 & -15 & -2.93 & -2.06 & 4.89 & 1.32 \\
        $f^2 h_0 \theta_0$ & 137 & -- & -3.15 & -134 & -- & -- & -- & -- & 0.812 & 7 & -3.06 & -- \\
    \end{tabular}}
    \caption{Step~2 predictor weights $\mathbf{G}^L$ for the lift and $\mathbf{G}^M$ for the moment with rows grouped as in table~\ref{tab:features}. A dash marks a feature dropped by the 95\% selection, whose weight is zero in the matrix.}
    \label{tab:coef_step2}
\end{center}
\end{table}
%

\subsection{Accuracy of prediction model}
\label{sec:error}

\begin{figure}
    \centering
    \begin{subfigure}[t]{0.34\textwidth}
      \caption{}\label{fig:8a}
      \includegraphics[width=\linewidth]{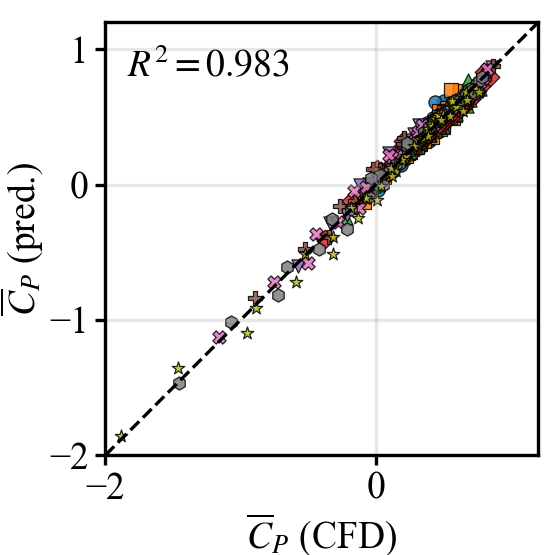}
    \end{subfigure}\hfill
    \begin{subfigure}[t]{0.64\textwidth}
      \caption{}\label{fig:8b}
      \includegraphics[width=\linewidth]{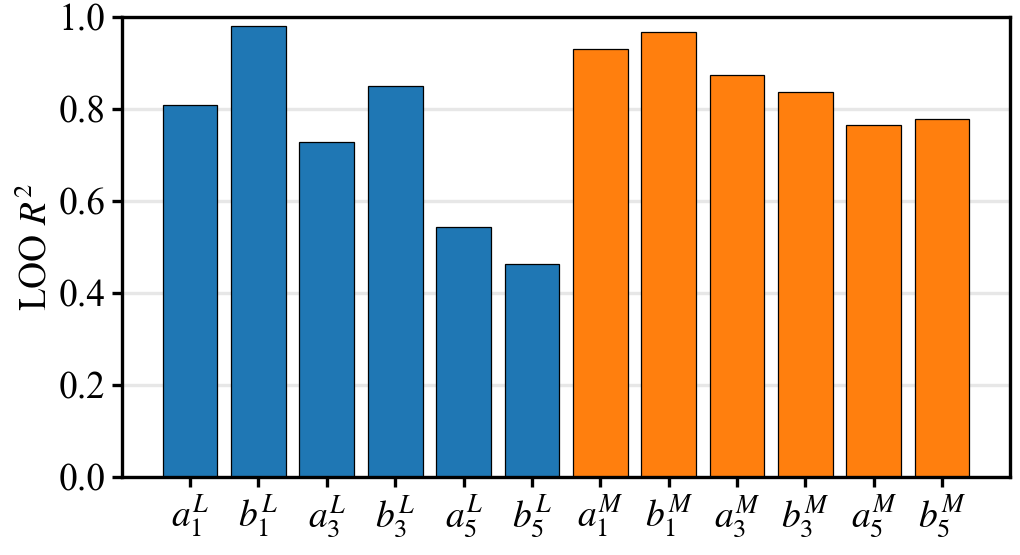}
    \end{subfigure}\\[4pt]
    \begin{subfigure}[t]{0.33\textwidth}
      \caption{}\label{fig:8c}
      \includegraphics[width=\linewidth]{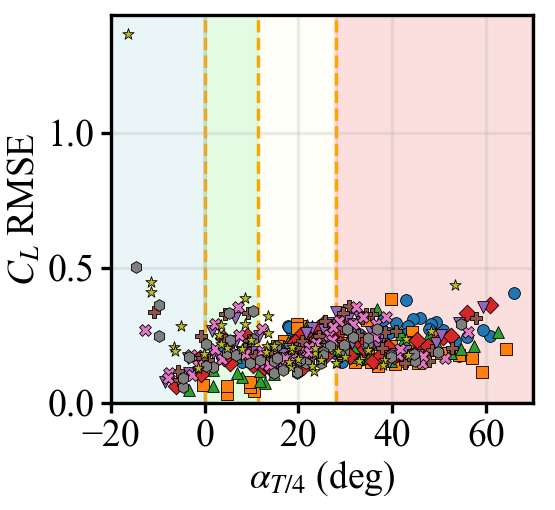}
    \end{subfigure}\hfill
    \begin{subfigure}[t]{0.33\textwidth}
      \caption{}\label{fig:8d}
      \includegraphics[width=\linewidth]{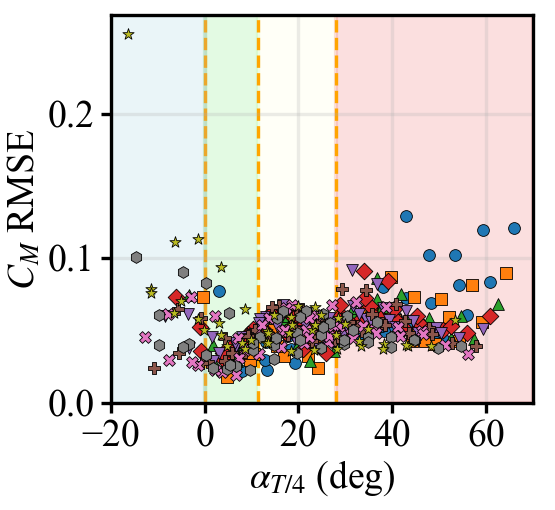}
    \end{subfigure}\hfill
    \begin{subfigure}[t]{0.33\textwidth}
      \caption{}\label{fig:8e}
      \includegraphics[width=\linewidth]{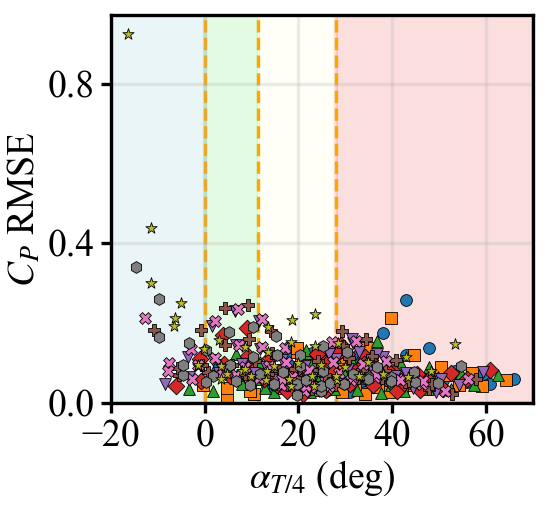}
    \end{subfigure}\\[4pt]
    \includegraphics[width=0.75\textwidth]{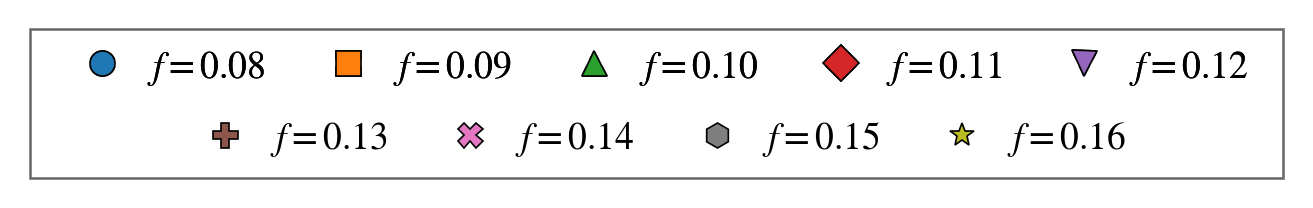}
    \caption{Prediction-error structure.
    (\textit{a}) Comparison between LOO predicted and ground truth for the cycle-averaged power $\bar{C}_P$. 
    (\textit{b}) LOO $R^2$ of the twelve Fourier coefficients. 
    (\textit{c}--\textit{e}) RMSE of $C_L$, $C_M$, and $C_P$ against $\alpha_{T/4}$, with the four flow regimes shaded as in figure~\ref{fig:distribution}. 
    }
    \label{fig:error_structure}
\end{figure}

The $\bar{C}_P$ parity plot (figure~\ref{fig:8a}) demonstrates the accuracy of the mean power prediction with LOO $R^2 = 0.98$ over all 405 samples. This accuracy comes from the two coefficients that enter $\bar{C}_P$ (\S\ref{sec:mean_power}), $b_1^L$ and $a_1^M$, which the Step~2 predictor recovers at $R^2 = 0.98$ and $0.93$ (figure~\ref{fig:8b}). The marginal drop from the Step~1 values ($0.99 \to 0.98$ for $b_1^L$) is due to the 95\% feature truncation between steps.

Figure~\ref{fig:8b} describes the accuracy of the remaining coefficients. In general there is a decline with harmonic order, with $k=1$ coefficients equal to or greater than $0.81$, but by $k=5$ the Fourier coefficients of the lift decrease to $0.54$ and $0.46$ for $a_5^L$ and $b_5^L$ respectively. The Fourier coefficients of the moment decrease less dramatically with respect to $k$, with a minimum of $0.77$ at $k=5$. These values closely mirror the trends and values of the Step~1 fits, since the truncation removes only the features that carry little weight. 

The difference in prediction quality is likely due to the different origins of the lift and moment. Although both integrate the same chord-wise pressure, the moment weights the pressure by the lever arm, which changes gradually with the kinematics and time, so its coefficients depend smoothly on the kinematics and fit better with the high-order content. However, the lift can change abruptly due to vortex shedding, a process captured by higher-order coefficients, and more difficult to fit.
%


\subsubsection{Error and accuracy in the time domain}

Whereas the Fourier-coefficient prediction is assessed with LOO $R^2$, the time-domain accuracy is measured by the leave-one-out root-mean-square error (LOO RMSE). The LOO RMSE of each sample is calculated as  
\begin{equation}
    \mathrm{RMSE}_i \;=\;\sqrt{\frac{1}{N_t}\sum_{n=1}^{N_t} \left(y_i(t_n) - \hat{y}_i(t_n)\right)^2},
    \label{eq:rmse}
\end{equation}
where $y_i(t_n)$ is the ground truth from simulation and $\hat{y}_i(t_n)$ its analytical prediction for simulation $i$ at the $n$th of $N_t$ samples over one cycle. The time-resolved lift and moment are reconstructed from the predicted Fourier coefficients, so the coefficient error is projected onto the instantaneous $C_L(t)$ and $C_M(t)$. Figures~\ref{fig:8c}--\ref{fig:8e} plot the RMSE of $C_L$, $C_M$, and $C_P$ against $\alpha_{T/4}$. The error remains bounded and nearly uniform across the parameter space, except for negative $\alpha_{T/4}$. The median RMSE is $0.20$ for $C_L$, $0.05$ for $C_M$, and $0.07$ for $C_P$, while the largest individual errors are confined to a few samples in the high-$h_0$ and low-$\theta_0$ (propulsion) region, where $C_L$ RMSE reaches $1.37$. Since the error does not grow with the response amplitude, the model reconstructs the strongest energy harvesting regions as accurately as the weakest ones.

\begin{figure}
    \centering
    \begin{subfigure}[t]{0.33\textwidth}\caption{\hfill$(0.50,\,45^\circ)$\hfill}\label{fig:9a}
      \includegraphics[width=\linewidth]{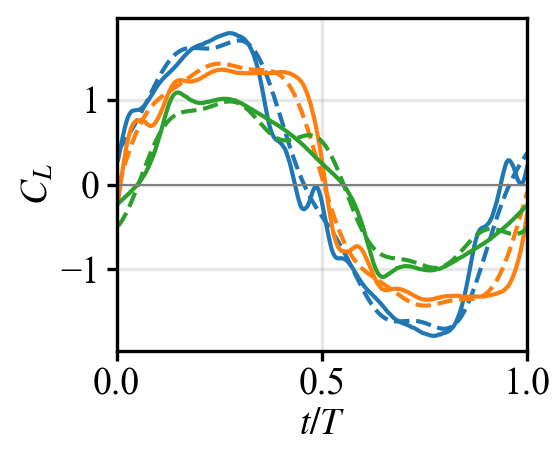}\end{subfigure}\hfill
    \begin{subfigure}[t]{0.33\textwidth}\caption{\hfill$(0.50,\,60^\circ)$\hfill}\label{fig:9b}
      \includegraphics[width=\linewidth]{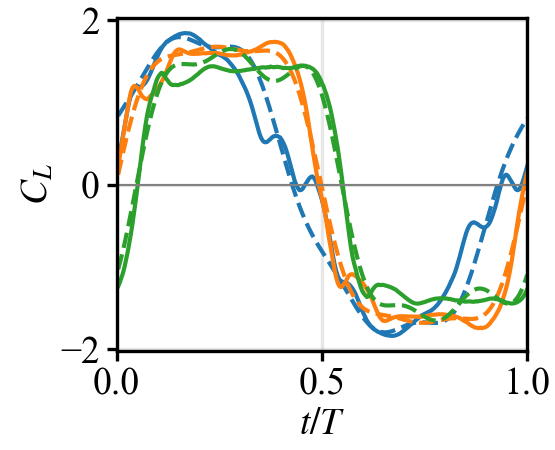}\end{subfigure}\hfill
    \begin{subfigure}[t]{0.33\textwidth}\caption{\hfill$(0.50,\,75^\circ)$\hfill}\label{fig:9c}
      \includegraphics[width=\linewidth]{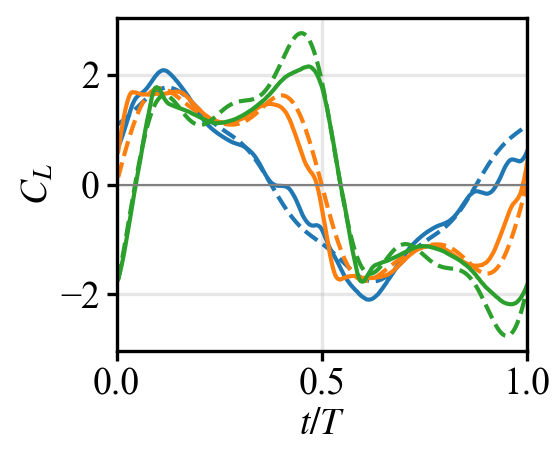}\end{subfigure}\\[3pt]
    \begin{subfigure}[t]{0.33\textwidth}\caption{\hfill$(1.00,\,45^\circ)$\hfill}\label{fig:9d}
      \includegraphics[width=\linewidth]{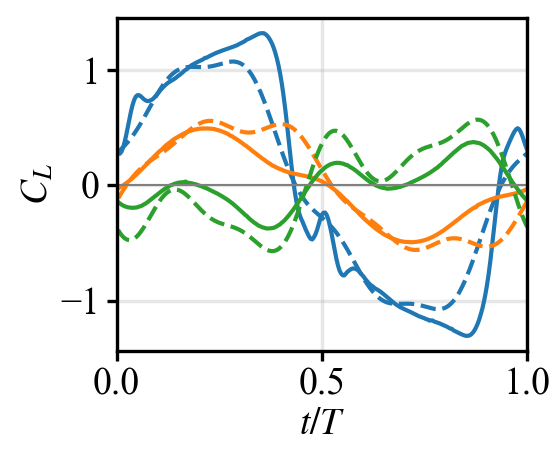}\end{subfigure}\hfill
    \begin{subfigure}[t]{0.33\textwidth}\caption{\hfill$(1.00,\,60^\circ)$\hfill}\label{fig:9e}
      \includegraphics[width=\linewidth]{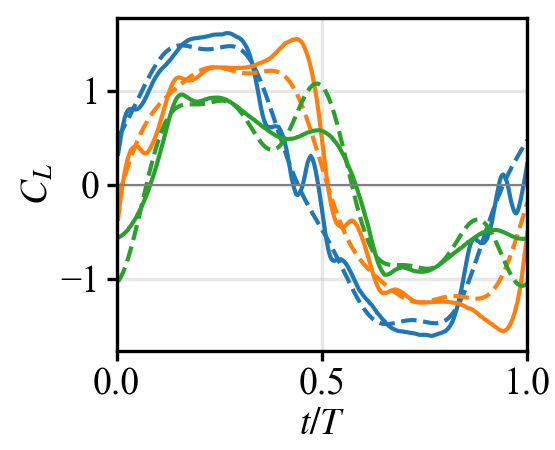}\end{subfigure}\hfill
    \begin{subfigure}[t]{0.33\textwidth}\caption{\hfill$(1.00,\,75^\circ)$\hfill}\label{fig:9f}
      \includegraphics[width=\linewidth]{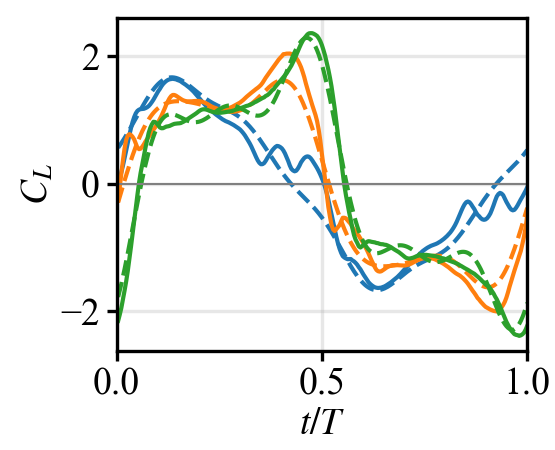}\end{subfigure}\\[3pt]
    \begin{subfigure}[t]{0.33\textwidth}\caption{\hfill$(1.50,\,45^\circ)$\hfill}\label{fig:9g}
      \includegraphics[width=\linewidth]{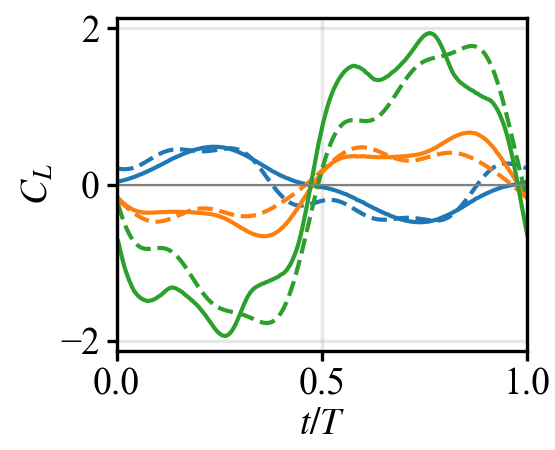}\end{subfigure}\hfill
    \begin{subfigure}[t]{0.33\textwidth}\caption{\hfill$(1.50,\,60^\circ)$\hfill}\label{fig:9h}
      \includegraphics[width=\linewidth]{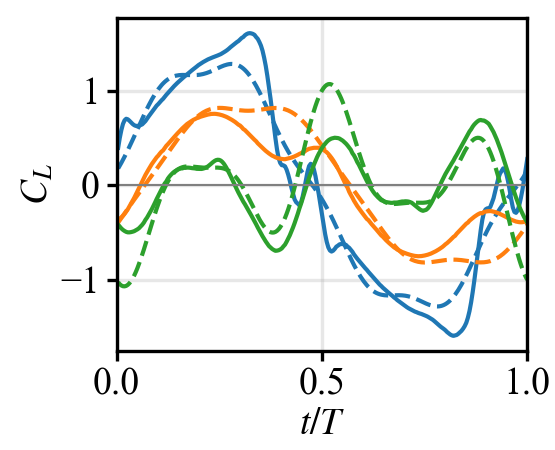}\end{subfigure}\hfill
    \begin{subfigure}[t]{0.33\textwidth}\caption{\hfill$(1.50,\,75^\circ)$\hfill}\label{fig:9i}
      \includegraphics[width=\linewidth]{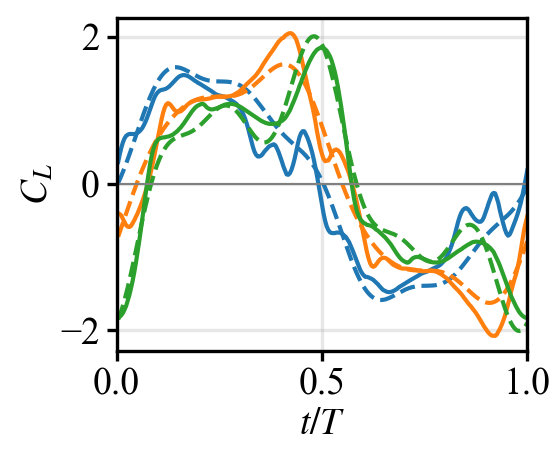}\end{subfigure}\\[3pt]
    \includegraphics[width=0.5\textwidth]{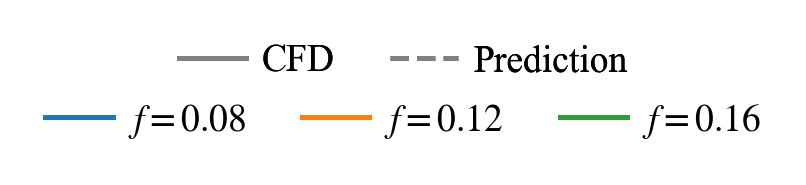}
    \caption{Simulation (solid) versus closed-form prediction (dashed) of $C_L(t)$ across the kinematic sweep. Each panel is labeled by its $(h_0,\theta_0)$, with $h_0$ increasing down the rows and $\theta_0$ across the columns. Within each panel, three frequencies $f = 0.08, 0.12, 0.16$ are overlaid by color.}
    \label{fig:waveform_sweep}
\end{figure}

Figure~\ref{fig:waveform_sweep} overlays the predicted and computed lift across a sample of the kinematic space. Similar analysis of the moment reconstruction is shown in appendix~\ref{sec:appendix_sweep}. The prediction follows the simulation throughout the parameter space. At low $\theta_0$ the lift is approximately sinusoidal and the two curves nearly coincide. As $\theta_0$ increases, the third and fifth harmonics grow and the lift develops multiple peaks. The prediction captures these peaks in both shape and phase. When $h_0$ and $\theta_0$ are both high, the prediction model reproduces the dominant peaks but smooths the small, high-frequency oscillations carried by $k>5$ harmonics. These departures are minor and confined to sharp peaks of the lift. 

For some kinematic combinations, the error is primarily in the peak height, not the timing. As $f$ increases, the lift contains more high-order content. The third and fifth harmonics increase from about $3\%$ of the spectral content at $f=0.08$ to about $15\%$ at $f=0.16$. At low $f$ they are too small to affect the absolute error of the peaks, while at high $f$ they set the peak shape, which shows in the discrepancy.

%


\subsubsection{Peak and phase error}

Since RMSE averages the point-wise error over the cycle, the peak lift and the timing of the peak lift are also analyzed. The peak error is the relative error of the peak lift, which is defined as
\begin{equation}
    \left|\delta C_{L,\max}\right|
      = \left|\frac{\hat{C}_{L,\max} - C_{L,\max}}{C_{L,\max}}\right|,
    \label{eq:peak_err}
\end{equation}
where $C_{L,\max}$ and $\hat{C}_{L,\max}$ are the simulated and predicted maximum lift over one cycle. Of the 357 energy-harvesting kinematic combinations ($\bar{C}_P>0$), the median peak error is $9.0\%$ (mean $12.6\%$). It grows with frequency, from about $7\%$ at $f=0.08$ to $13\%$ at $f=0.16$. The peak is sharpened by the third and fifth harmonics, which are predicted least well (\S\ref{sec:step1_ridge}) and are most important at high $f$.

The phase error $\tau$ is the lag $s$ that maximizes the periodic cross-correlation of the predicted and simulated responses, normalized by the period $T$,
\begin{equation}
    \tau = \frac{1}{T}\,\operatorname*{arg\,max}_{s}\,\int_0^T C_L(t)\,\hat{C}_L(t+s)\,\mathrm{d}t.
\end{equation}
%
The median $\tau$ is $0.008$, and $87\%$ of the kinematic combinations lie within $0.02$. Unlike the peak, $\tau$ shows no dependence on frequency. The moment has lower phase error than lift, with details given in appendix~\ref{sec:appendix_sweep}.

Figure~\ref{fig:peak_phase_error} shows peak and phase error for lift as a function of $\alpha_{T/4}$.  Both errors grow sharply at small $\alpha_{T/4}$, where twenty kinematic combinations carry a phase error above $0.05$ and one peak error reaches $234\%$. These combinations generate almost no power, and their Fourier coefficients are correspondingly small and harder to predict accurately. The reconstructed lift therefore departs from the simulation and drifts out of phase with it. These peak and phase errors are prominent for regions unlikely to be utilized for energy harvesting, since the efficiency and power generation are very low. 

\begin{figure}
    \centering
    \begin{subfigure}[t]{0.5\textwidth}\caption{}\label{fig:10a}
      \includegraphics[width=\linewidth]{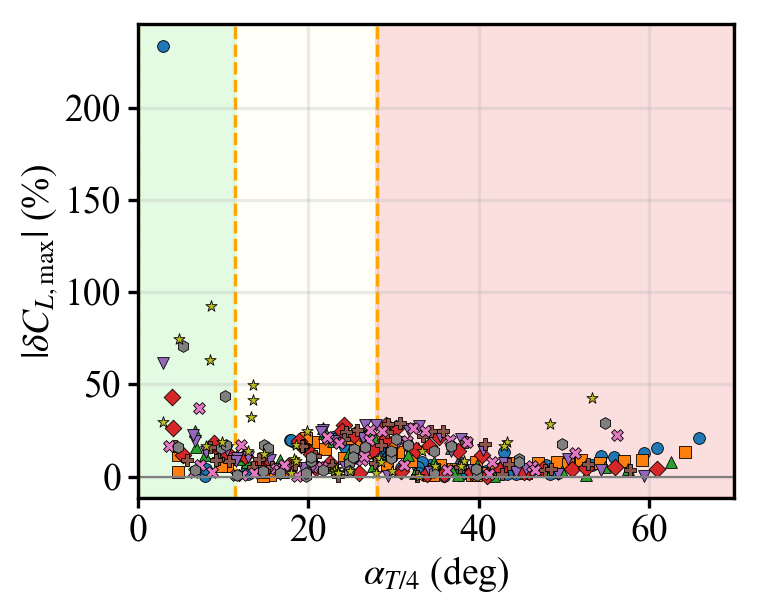}\end{subfigure}\hfill
    \begin{subfigure}[t]{0.5\textwidth}\caption{}\label{fig:10b}
      \includegraphics[width=\linewidth]{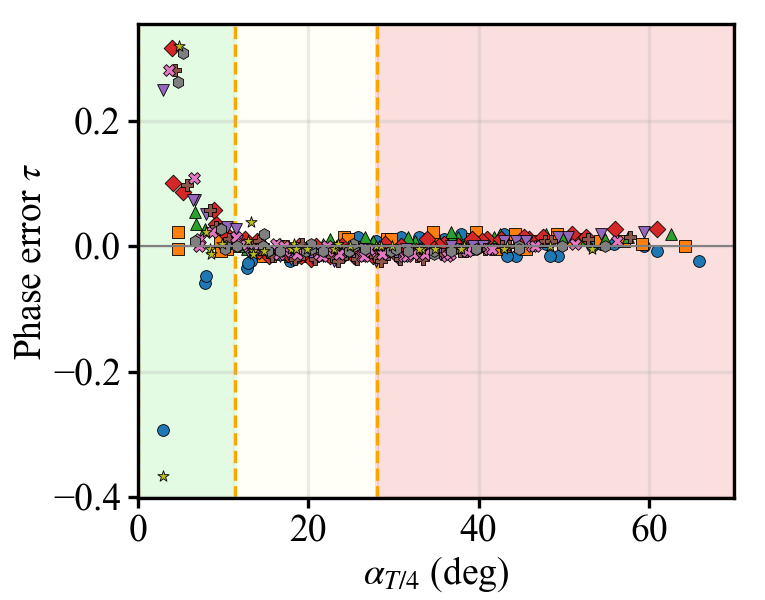}\end{subfigure}\\[3pt]
    \includegraphics[width=0.75\textwidth]{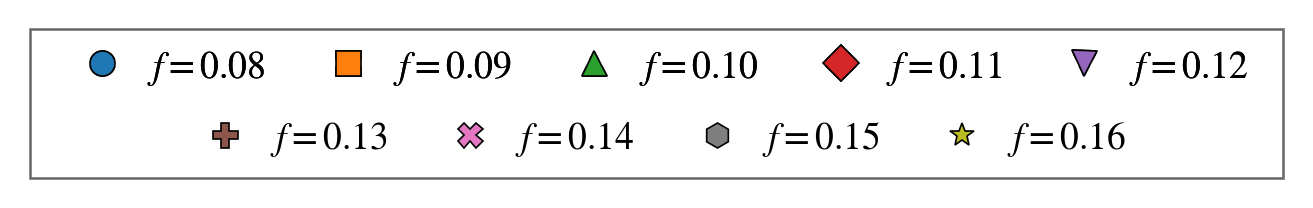}
    \caption{Prediction accuracy of (\textit{a}) the peak lift magnitude (relative error) and (\textit{b}) the lift phase error (full-cycle cross-correlation, as a fraction of the period) versus $\alpha_{T/4}$, for all energy-harvesting kinematic combinations ($\bar{C}_P > 0$).}
    \label{fig:peak_phase_error}
\end{figure}
%


\section{Discussion}
\label{sec:shedding}

\subsection{Vortex shedding content in high-order harmonics}
\label{sec:events}

This section connects the harmonic content directly to the vortex events in the flow, such as the formation and shedding of a coherent vortex. When the boundary layer remains attached, the lift varies smoothly at the stroke frequency and is well approximated by the $k=1$ harmonic. A single leading-edge vortex adds one localized force excursion per half-cycle, which is captured by the $k=3$ harmonic. If the stroke sheds a leading-edge and then a trailing-edge vortex, the two excursions activate the $k=5$ harmonic. The dominant harmonic above $k=1$ therefore counts the discrete vortex events per stroke.

\begin{figure}
    \centering
    \begin{subfigure}{\textwidth}\caption{~$f=0.08$, $h_0=1.25$, $\theta_0=40^\circ$}\label{fig:11a}
      \includegraphics[width=\linewidth]{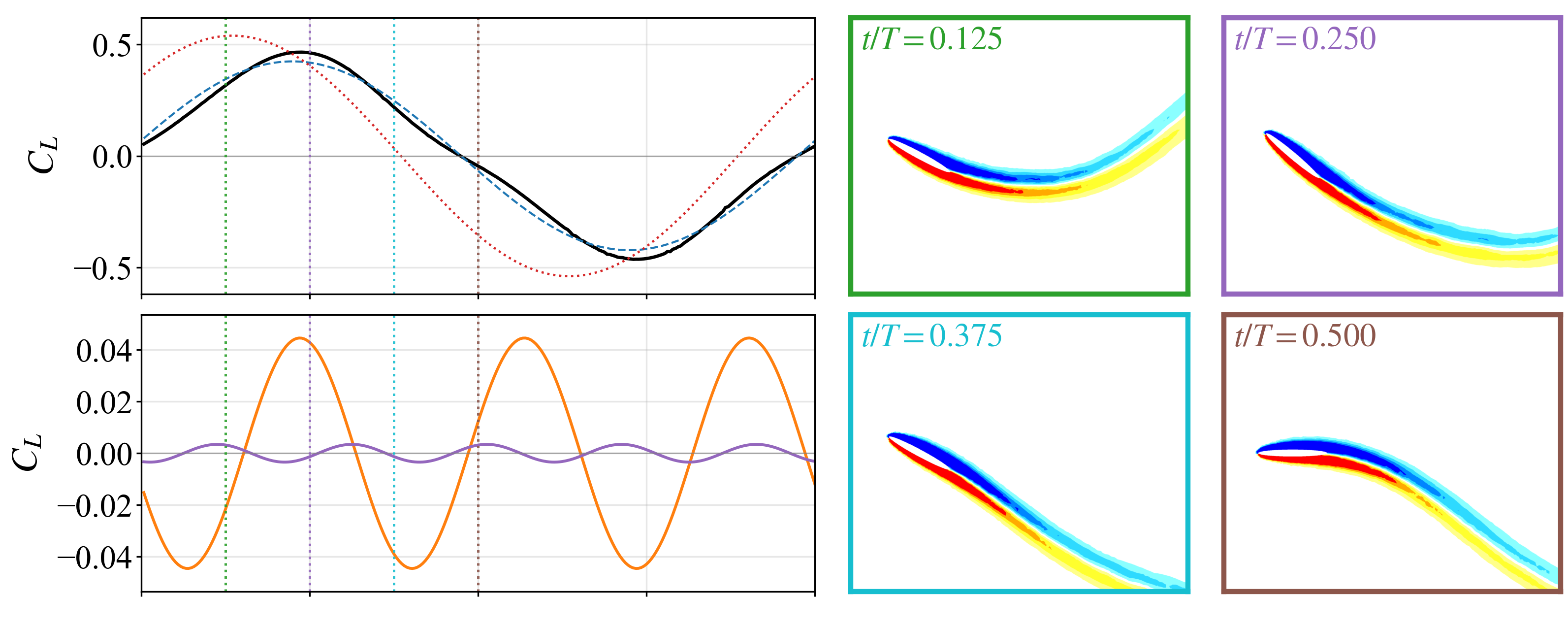}\end{subfigure}
    \begin{subfigure}{\textwidth}\caption{~$f=0.09$, $h_0=1.00$, $\theta_0=40^\circ$}\label{fig:11b}
      \includegraphics[width=\linewidth]{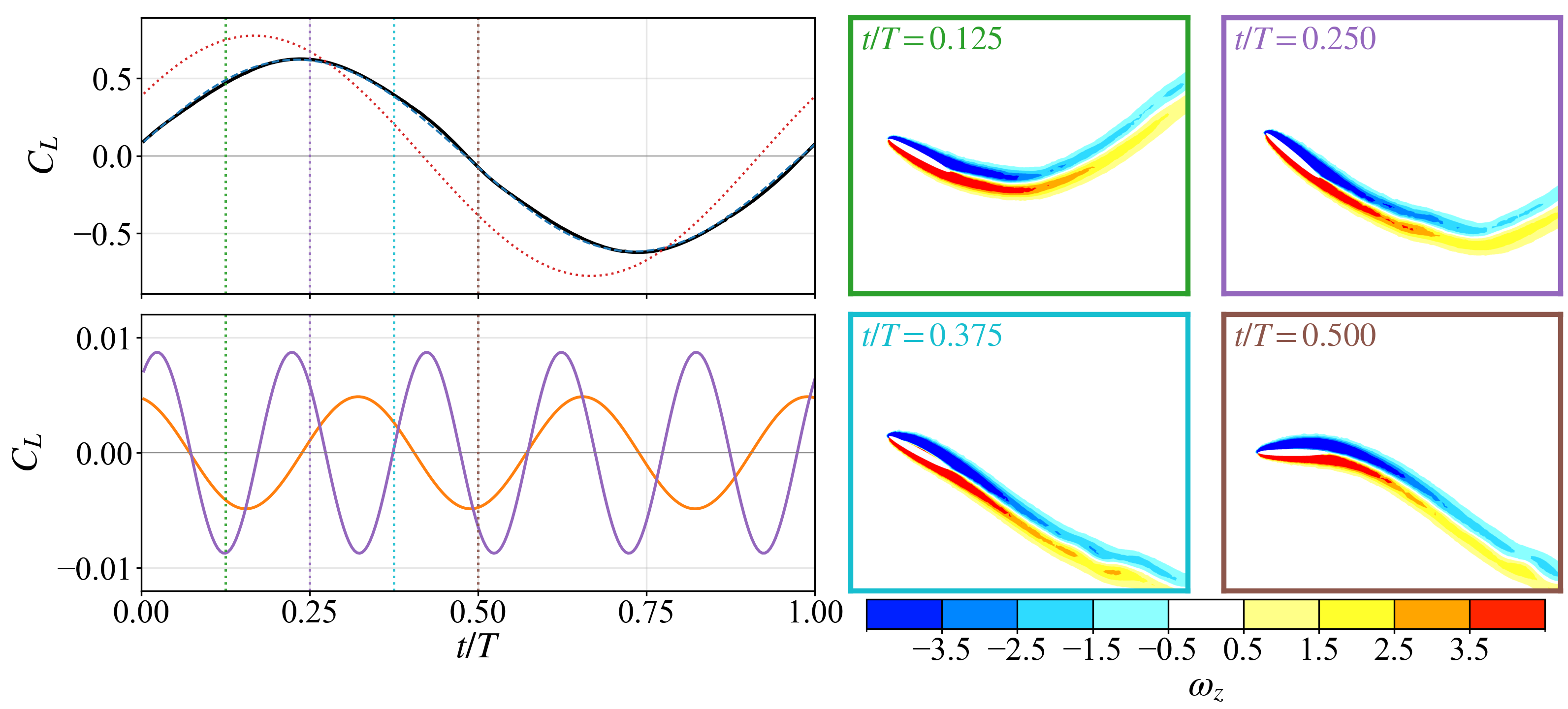}\end{subfigure}
    \includegraphics[width=\linewidth]{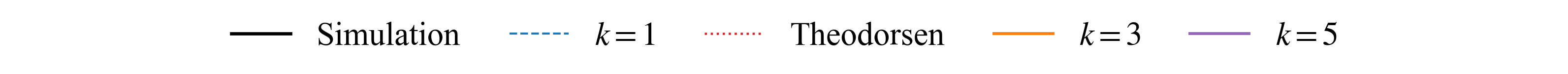}
    \caption{Harmonic--event correspondence for two near-sinusoidal ($k=1$ dominated) examples. In each panel, the upper plot shows $C_{L}(t)$ from the simulation (solid black), its $k=1$ component (dashed blue), and Theodorsen's prediction (dotted red). The lower plot shows the $k=3$ (orange) and $k=5$ (purple) components on a separate scale. The four snapshots show the vorticity at $t/T = 0.125, 0.250, 0.375, 0.500$, the instants marked by the vertical lines. The snapshots cover the upstroke half-cycle, and the second half repeats them with opposite sign.}
    \label{fig:mode_event_k1}
\end{figure}

\begin{figure}
    \centering
    \begin{subfigure}{\textwidth}\caption{~$f=0.16$, $h_0=0.50$, $\theta_0=80^\circ$}\label{fig:12a}
        \includegraphics[width=\linewidth]{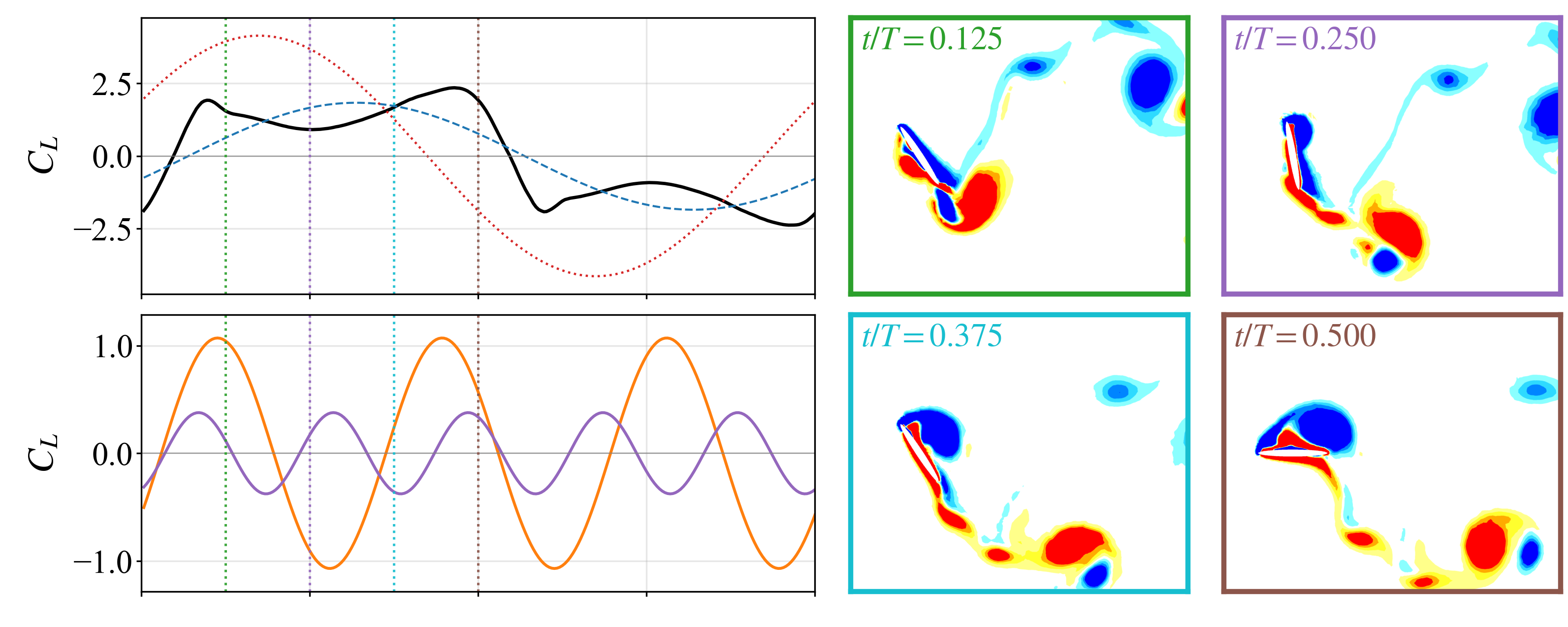}\end{subfigure}
    \begin{subfigure}{\textwidth}\caption{~$f=0.13$, $h_0=1.00$, $\theta_0=80^\circ$}\label{fig:12b}
        \includegraphics[width=\linewidth]{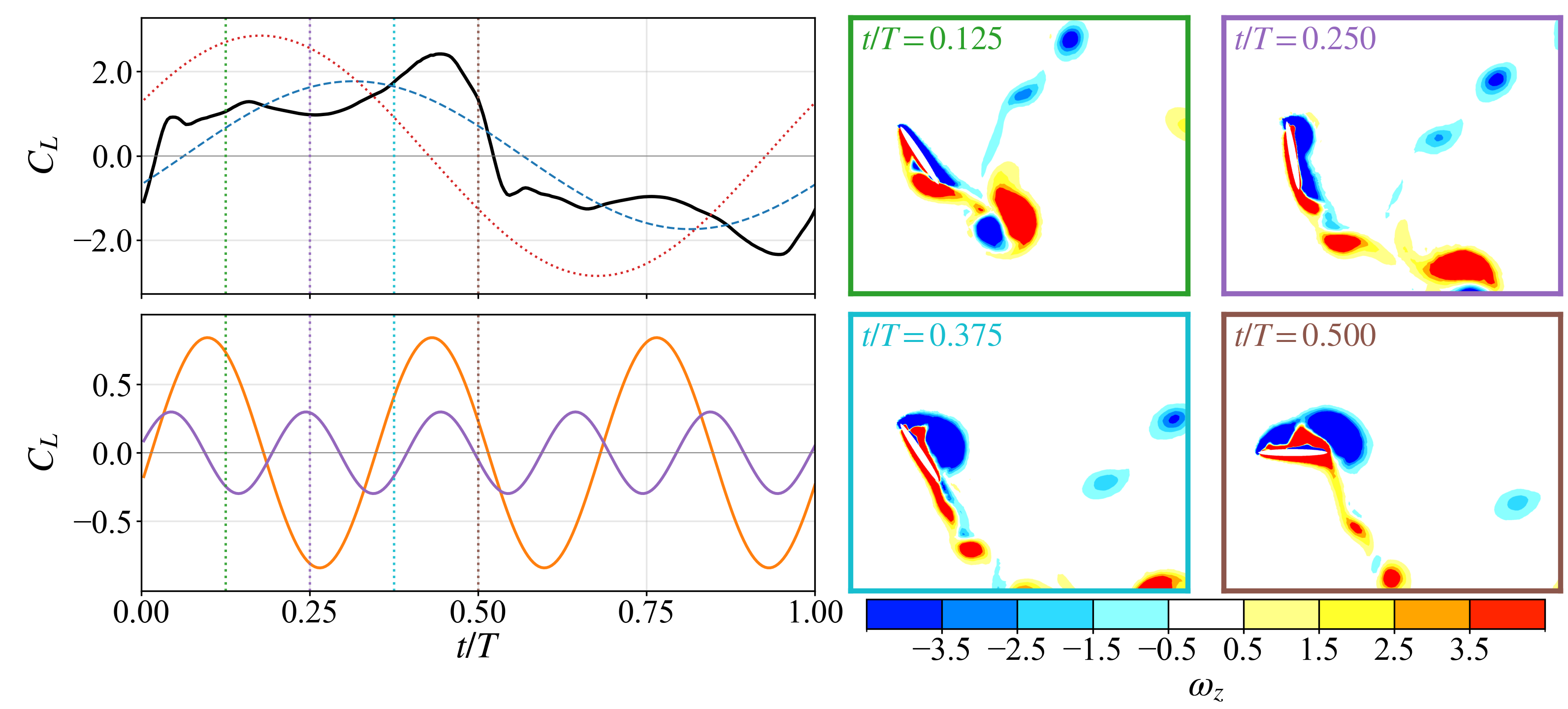}\end{subfigure}
    \includegraphics[width=\linewidth]{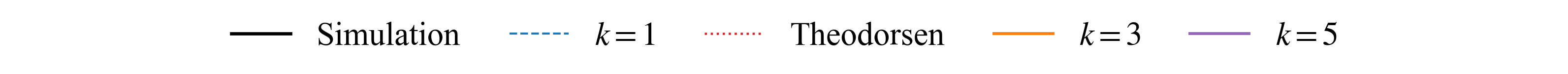}
    \caption{Same as figure~\ref{fig:mode_event_k1}, for two examples in which the third harmonic leads the high-order content.}
    \label{fig:mode_event_k3}
\end{figure}

\begin{figure}
    \centering
    \begin{subfigure}{\textwidth}\caption{~$f=0.09$, $h_0=0.50$, $\theta_0=60^\circ$}\label{fig:13a}
      \includegraphics[width=\linewidth]{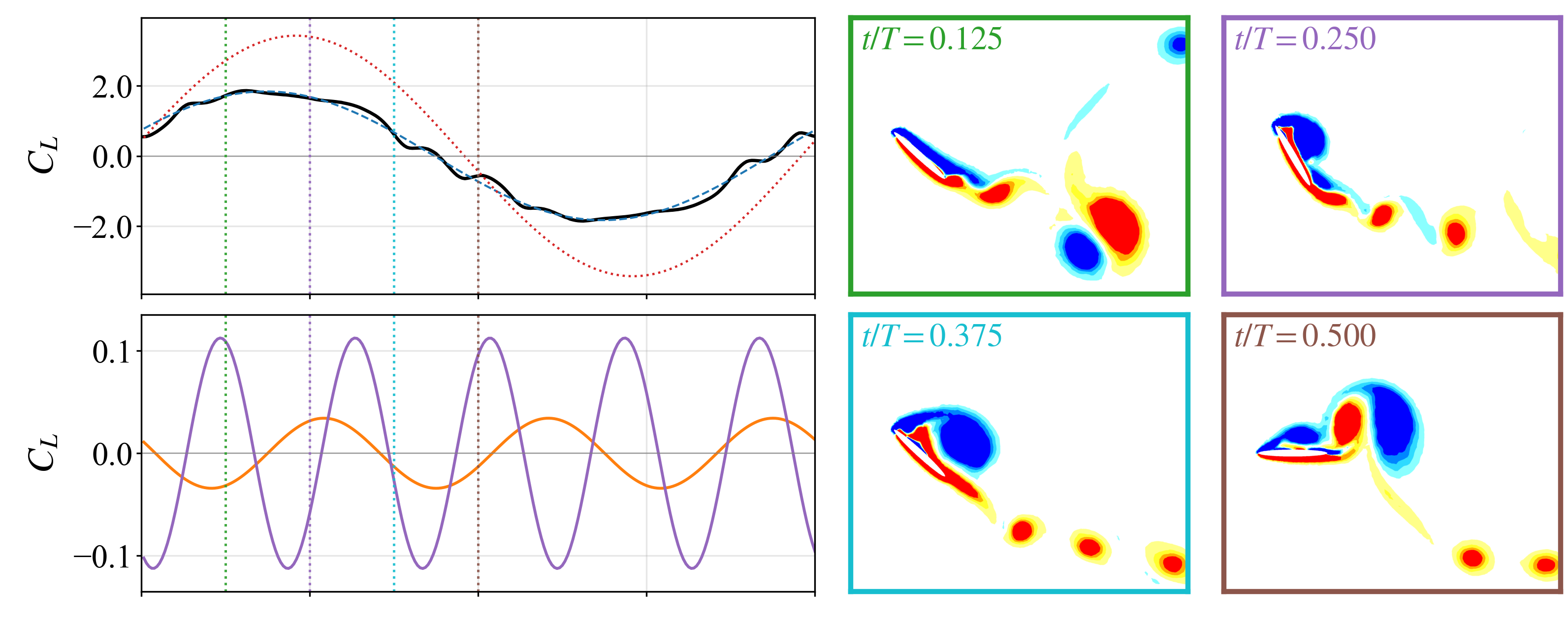}\end{subfigure}
    \begin{subfigure}{\textwidth}\caption{~$f=0.08$, $h_0=1.00$, $\theta_0=60^\circ$}\label{fig:13b}
      \includegraphics[width=\linewidth]{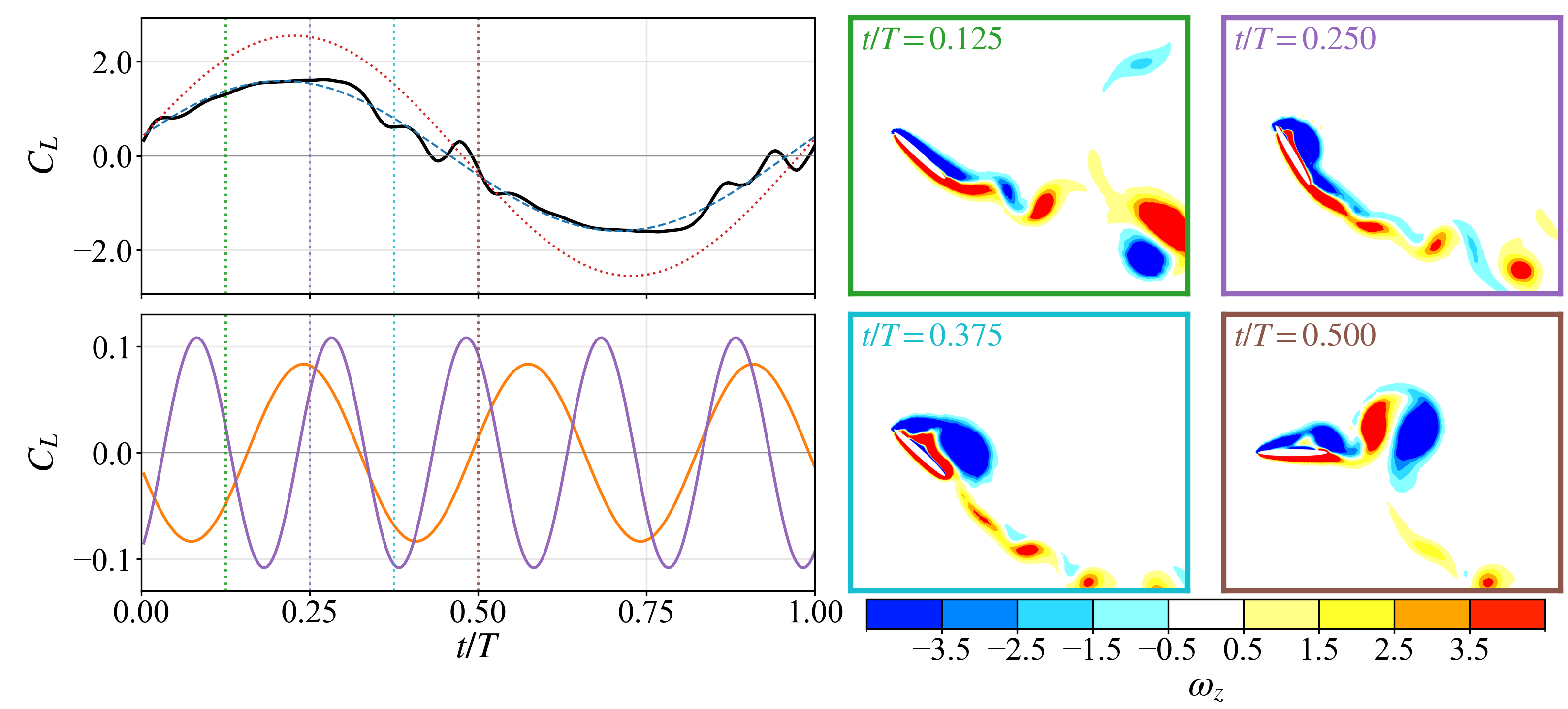}\end{subfigure}
    \includegraphics[width=\linewidth]{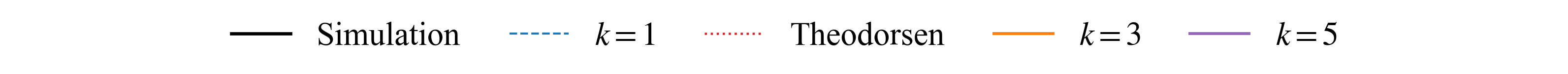}
    \caption{Same as figure~\ref{fig:mode_event_k1}, for two examples in which the fifth harmonic leads the high-order content.}
    \label{fig:mode_event_k5}
\end{figure}

Figures~\ref{fig:mode_event_k1}--\ref{fig:mode_event_k5} show six examples spanning the observed range of the high-order content. Each panel is one set of oscillation kinematics. The upper plot shows the lift, its $k=1$ component, and the classical unsteady aerodynamics theory of \citet{Theodorsen1935} as a reference, and the lower plot shows the $k=3$ and $k=5$ components. The four snapshots show the vorticity field at $t/T = 0.125$, $0.25$, $0.375$, and $0.5$, the instants marked by the vertical lines in the plots.

In figure~\ref{fig:mode_event_k1}, the representative examples have low frequencies and small pitch amplitude. The boundary layer is attached to the foil and develops into a smooth, thin wake without any vortex formation. Zero discrete events result in little to no high-order content, as the simulated lift remains on its $k=1$ component and the high-order harmonics vanish. The prediction by Theodorsen does not match the $k=1$ component of the simulation, leading it in phase and over-predicting the amplitude. 

%
%

In the examples of figure~\ref{fig:mode_event_k3}, a leading-edge vortex now begins to form early in the stroke and grows as it moves along the chord. The lift tracks this single event, exceeding the $k=1$ component early in the stroke and passing through a local minimum near the mid-stroke, where the leading-edge vorticity has not yet rolled up. The vortex grows and rolls up in the second half of the stroke, and the pitch rotation retains it on the surface. The lift therefore reaches a second local peak and remains above the $k=1$ component until the reversal, where the vortex sheds and the lift decreases steeply.  This sequence requires an incidence large enough to separate the leading edge into a coherent vortex and a stroke period comparable with the formation time of the vortex, so that the reversal arrives before the vortex would pinch off \citep{Gharibetal1998, EldredgeJones2019}, a condition satisfied with large pitch amplitudes ($\theta_0 \geq 65^\circ$) and high frequencies ($0.13\leq f \leq 0.16$). Each stroke contains two peaks in the lift, and the third harmonic dominates the high-order content with an amplitude half of the first harmonic and three times that of the fifth.  
%

%

The fifth harmonic leads the high-order content when $f=0.08$--$0.09$ and $\theta_0=60$--$75^\circ$, as demonstrated in figure~\ref{fig:mode_event_k5}. A leading-edge vortex forms early in the heave stroke. The long half-stroke exceeds the timescale of LEV formation, so the vortex pinches off before stroke reversal. The foil continues to heave and forms a trailing-edge vortex from the pressure shear layer. The two counter-rotating vortices form a pair at reversal, and a secondary LEV forms during the pitch reversal. These events activate the fifth harmonic of the lift, with an amplitude larger than that of the third harmonic and about $7\%$ of the first harmonic. The high-order content of both examples in figure~\ref{fig:mode_event_k5} is overall weaker than that of the $k=3$ examples (figure~\ref{fig:mode_event_k3}), thus producing smaller amplitude fluctuations rather than the two distinct peaks of a single-LEV stroke in figure~\ref{fig:mode_event_k3}.
%

%

Although not shown, the vortex-shedding events also affect the moment. The moment follows the position of the vortex, and the primary LEV moves along the chord once per stroke. In all six examples shown in figures~\ref{fig:mode_event_k1}--\ref{fig:mode_event_k5}, the moment's high-order content is led by the third harmonic, even when the lift has significant fifth-harmonic contributions. For all 405 simulations in figure~\ref{fig:5b}, the first and third harmonics contain at least $95\%$ of the moment's spectral content.  

Since the number of events per cycle depends on how the stroke half-period compares with the vortex formation time, it is primarily controlled by the frequency of the foil. Across the $a$ and $b$ coefficients, $f$ is a leading contribution at every harmonic in figure~\ref{fig:ridge_heatmap} (\S\ref{sec:step1_ridge}). The vortex convects along the chord over a time set mainly by the free stream, so a change in $f$ shifts when the events occur within the cycle. A shift $\Delta t$ rotates the $k$th harmonic by $2\pi k\,\Delta t/T$, five times as far at $k=5$ as at $k=1$. The high harmonics therefore respond to frequency nonlinearly, and their attribution concentrates increasingly on $f^2$ and $f^2\theta_0$. The same sensitivity makes the $k=5$ coefficients of the lift the weakest targets in both Step~1 (\S\ref{sec:step1_ridge}) and Step~2 (\S\ref{sec:error}). With the same sixteen features, a Step~1 fit to the amplitude $A_5$ is $R^2 = 0.76$, compared with $0.64$ and $0.46$ for its components $a_5^L$ and $b_5^L$. However, it is still worth including the fifth harmonic in the overall prediction since it is the lowest-order harmonic that can represent two events in a stroke, the LEV and the TEV.


\subsection{Comparison with Theodorsen's model}
\label{sec:theodorsen}

As a reference point for classical unsteady aerodynamic approximations, figures~\ref{fig:mode_event_k1}--\ref{fig:mode_event_k5} compared the simulation results and their first harmonics to the Theodorsen model of $C_L^{T}(t)$, 
\begin{equation}\label{eq:theodorsen_lift}
  C_L^{T}(t) = \underbrace{2\pi \,\mathcal{C}(\pi f)\big[\,\theta + \dot h + \tfrac{1}{4}\,\dot\theta\,\big]}_{\text{circulatory}} + \underbrace{\tfrac{1}{2}\pi \big(\ddot h + \dot\theta\big)}_{\text{added mass}},
\end{equation}
which calculates the lift coefficient from the pitch, velocity, and acceleration terms. 
The Theodorsen function, $\mathcal{C}$, accounts for the influence of the shed wake and is evaluated at the reduced frequency $\pi f$, 
\begin{equation}\label{eq:theodorsen_C}
  \mathcal{C}(\pi f) = \frac{H_1^{(2)}(\pi f)}{H_1^{(2)}(\pi f) + \mathrm{i}\,H_0^{(2)}(\pi f)},
\end{equation}
where $H_n^{(2)}$ is the Hankel function of the second kind of order $n$. The magnitude of $\mathcal{C}$ scales the circulatory lift and the argument of $\mathcal{C}$ introduces a phase lag. The input kinematics ($\theta$, $\dot h$, $\dot\theta$, and $\ddot h$) correspond to the features ($\theta_0$, $fh_0$, $f\theta_0$, and $f^2h_0$), but one must also account for the frequency dependence in $\mathcal{C}(\pi f)$. From $f=0.08$ to $0.16$ the magnitude of $\mathcal{C}$ decreases from $0.72$ to $0.62$, while its phase lag varies between $14^\circ$ and $15^\circ$. A linear fit $\mathcal{C}(\pi f) \approx c_0 + c_1 f$, with complex constants $c_0 = 0.775 - 0.222\mathrm{i}$ and $c_1 = -1.154 + 0.446\mathrm{i}$, reproduces $\mathcal{C}(\pi f)$ to within $1.3\%$ over this range of $f$. Adding a linear dependence on $f$ to the circulatory terms reveals five features ($\theta_0$, $fh_0$, $f\theta_0$, $f^2h_0$ and $f^2\theta_0$) that are accounted for in Theodorsen's approximation. In figure~\ref{fig:ridge_heatmap}, these five features occupy only three of the six groups and contain $30.9\%$ of the $a_1^L$ attribution and $40.4\%$ of $b_1^L$. All five are first order in the motion amplitudes, whereas much of the remaining attribution concentrates on cross-product terms such as $h_0\theta_0$ and $f\theta_0^2$ that are not present in Theodorsen's formulation. However, these missing terms are negligible at small $h_0$ and $\theta_0$, which is consistent with the small angle approximations of the theory. 

Theodorsen's model assumes an inviscid, attached flow past a thin foil in small-amplitude motion, with the shed vorticity convected downstream in a planar wake, so that a sinusoidal motion produces a sinusoidal lift without higher harmonics. The attached flows of figures~\ref{fig:11a} and~\ref{fig:11b} are the closest in the dataset to what the theory assumes, but the Theodorsen amplitude is $25\%$ above the simulation, and the phase is $24^\circ$ to $33^\circ$ offset. The amplitude mismatch is likely explained by the $\Rey = 1000$ flow and smaller lift slope, not satisfying the inviscid assumption of Theodorsen. 
Similarly, in figures~\ref{fig:mode_event_k3} and~\ref{fig:mode_event_k5} the amplitude of the first harmonic is overpredicted by a factor of $1.6$ to $2.3$.
The phase of the prediction is determined by the constant coefficients in equation~\eqref{eq:theodorsen_lift} and by the wake function $\mathcal{C}$. The lag of $\mathcal{C}$ is a nearly constant $15^\circ$ for all kinematics considered in this study. The remaining phase discrepancy may partly reflect missing kinematic features that depend on $h_0$ or $\theta_0$, as mentioned in the paragraph above, which are not negligible for the energy harvesting kinematics. 
\section{Conclusions}
\label{sec:discussion}

This research builds an analytical model to predict the instantaneous lift, pitching moment and power of an oscillating-foil energy harvester from its kinematics. The lift and moment are decomposed into Fourier harmonics, and each coefficient is mapped from the kinematics by Ridge regression on physics-derived features. Unlike state-based models, which map the instantaneous kinematic state to the instantaneous force, the model maps the three amplitude parameters to the whole cycle at once, and one fit covers the full kinematic sweep. The cycle-averaged power is recovered accurately (LOO $R^2 = 0.98$), and is computed from two Fourier coefficients, $b_1^L$ and $a_1^M$. Both belong to the first harmonic and are driven by the amplitude of the motion rather than by high-order cross-product terms. The instantaneous response is harder to predict, though the median peak error is $9.0\%$ and the median phase error is $0.008\, T$. Apart from a small residual left by the truncation at $k=5$, the error concentrates in the high harmonics, whose coefficients capture the timing of the vortex events. The remaining error reflects the amount of data rather than the choice of features in \S\ref{sec:feature}. A richer polynomial basis would not remove the error, since the 405 simulations leave only about twenty-five samples per feature and every added feature makes the weights less determined.

Due to the Fourier decomposition, the harmonic order can be used to count the discrete vortex events within a stroke. Predicting the Fourier coefficients from the kinematics is therefore the same as predicting how many vortices a stroke sheds and how strongly. When the flow remains attached throughout the stroke, the lift is dominated by the first harmonic. When a single LEV is shed per half cycle, the third harmonic is activated, and when the LEV pairs with a TEV the fifth harmonic dominates the high-order content. 
The vortex analysis is consistent with the attribution of the features, since the linear group dominates the first-harmonic coefficients while $f^2$ and $f^2\theta_0$ increase for the third and fifth harmonics. As a comparison, the classical unsteady theory of Theodorsen spans only five of the sixteen features, and cannot accurately recover the first harmonic, in either amplitude or phase. 

The scope of these results is relevant to the energy-harvesting regime of oscillating foils. The model is trained at a single Reynolds number, in two dimensions, for a $10\%$ elliptical foil pitching about its mid-chord at a fixed $90^\circ$ phase between heave and pitch. The prediction framework can be extended beyond the trained configuration. For example, other thin airfoils are expected to perform similarly, and pitching about any other chord location can be modeled by a change of reference point. However, a new fit would be required for a heave--pitch phase different from $90^\circ$. The prediction is weakest as the mean power approaches zero, but these kinematic conditions extract very little power, and are typically outside the operating range of an energy harvester.

The impacts of this model can be applied to the energy harvesting community in terms of a reduced-order model of power, forces, and torque on an oscillating foil, reducing the need for expensive parameter sweeps in simulations. Furthermore, the proposed model can be integral to more elaborate wake-foil interaction models, since it contains time-dependent information about how and when the power is generated throughout the stroke. 



\backsection[Acknowledegments]{The research was conducted using computational resources and services at the Center for High Throughput Computing (CHTC) at the University of Wisconsin--Madison.}

\backsection[Funding]{This research is supported by the National Science Foundation under award 2338957.}

\backsection[Declaration of interests]{The authors report no conflict of interest.}

\backsection[Data availability statement]{The data will be made available upon request.}

\appendix

\section{Drag prediction}\label{sec:appendix_drag}

The lift and moment fix the extracted power, but the drag is also predicted. It differs from the lift and moment, since the foil interacts with the oncoming flow the same way on the up- and down-strokes in the streamwise direction ($C_D(t+T/2)=C_D(t)$). Its Fourier series therefore contains only the even harmonics $k=0,2,4,\dots$ and it has a non-zero mean $a_0^{D}=\bar{C}_D$, which is the cycle-averaged drag. The truncation at $k=4$ retains at least 95\% of the spectral content in all but one of the 405 simulations, the exception reaching 91.6\% at the highest frequency and heave amplitude. The sixteen features, the two-step Ridge, and the closed-form predictor of \S\ref{sec:ridge} are otherwise unchanged. Table~\ref{tab:coef_D} shows all the coefficients for the drag prediction model.
%
%

%
\begin{table}
\begin{center}
    \renewcommand{\arraystretch}{1.2}
    \begin{tabular}{lccccc}
        \hline
        Feature & $a_0^D$ & $a_2^D$ & $b_2^D$ & $a_4^D$ & $b_4^D$ \\
        \hline
        $f$ & 17.2 & -1.6 & 38.6 & -5.36 & -2.96 \\
        $h_0$ & -0.816 & -0.338 & 2.9 & -2.6 & -0.551 \\
        $\theta_0$ & -0.0733 & -1.43 & -6.96 & 0.0956 & 1.2 \\
        \hline
        $f^2$ & -78.1 & -29.6 & -343 & 31.6 & -- \\
        $\theta_0^2$ & -- & -- & 8.49 & -0.424 & -1.72 \\
        \hline
        $f h_0$ & -9.03 & 22.4 & -22 & 54.1 & -0.0783 \\
        $f^2 h_0$ & -- & -- & -- & -268 & 47.7 \\
        $f h_0^2$ & -- & -- & 0.763 & -- & -- \\
        $f^2 h_0^2$ & -- & -- & -- & -- & -- \\
        \hline
        $f \theta_0$ & -0.498 & 8.21 & 10.1 & -- & -3.31 \\
        $f^2 \theta_0$ & -7.23 & -- & 447 & 19.7 & -5.44 \\
        $f \theta_0^2$ & 6.66 & -- & -72.7 & 11.3 & 15.4 \\
        $f^2 \theta_0^2$ & -- & -0.096 & -- & -83.6 & -30.5 \\
        \hline
        $h_0 \theta_0$ & 1.34 & 0.0911 & -3.5 & 3.2 & 1.46 \\
        \hline
        $f h_0 \theta_0$ & -- & -18.5 & 26.9 & -65.2 & -14.2 \\
        $f^2 h_0 \theta_0$ & -- & 6.42 & -16 & 324 & -- \\
        \hline
    \end{tabular}
    \caption{Step~2 predictor weights $\mathbf{G}^D$ for the drag with rows grouped as in table~\ref{tab:features}. A dash marks a feature dropped by the 95\% selection, whose weight is zero in the matrix.}
    \label{tab:coef_D}
\end{center}
\end{table}

\begin{figure}
  \centering
  \begin{subfigure}[t]{0.33\textwidth}
    \caption{}\label{fig:appD_err_a}
    \includegraphics[width=\linewidth]{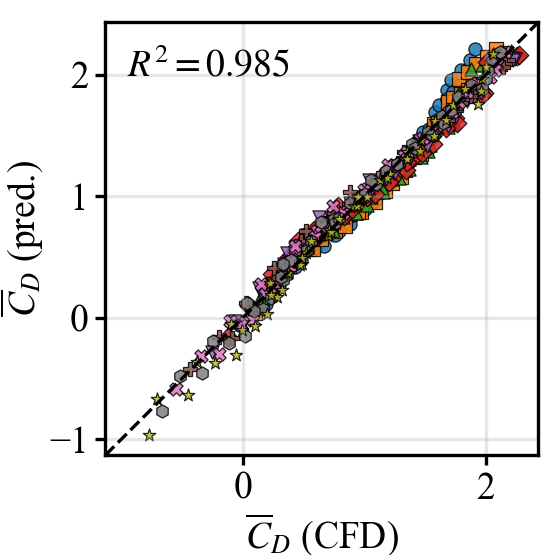}
  \end{subfigure}\hspace{2pt}
  \begin{subfigure}[t]{0.33\textwidth}
    \caption{}\label{fig:appD_err_b}
    \includegraphics[width=\linewidth]{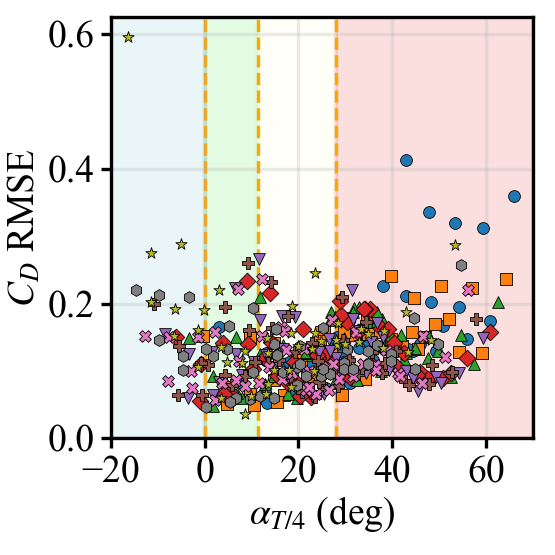}
  \end{subfigure}\\[4pt]
  \begin{subfigure}[t]{0.66\textwidth}
    \caption{}\label{fig:appD_err_c}
    \includegraphics[width=\linewidth]{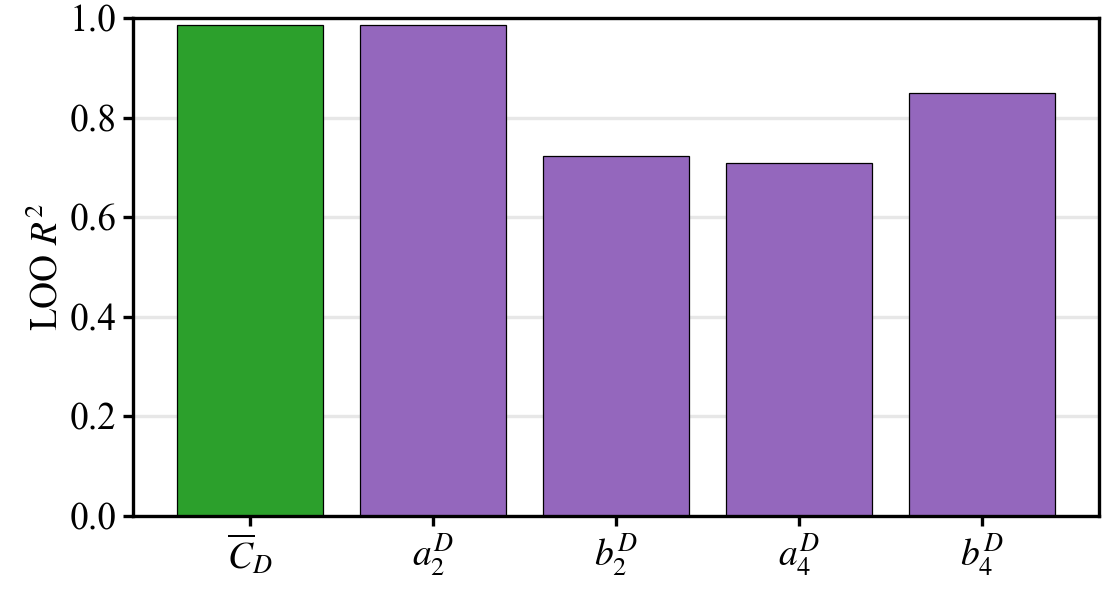}
  \end{subfigure}\\[4pt]
  \includegraphics[width=0.66\textwidth]{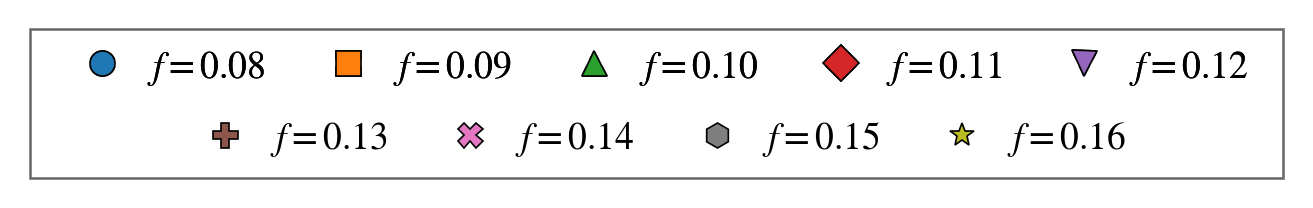}
  \caption{Prediction skill for the drag. (\textit{a}) Predicted against simulated mean
  drag $\bar{C}_D$, (\textit{b}) the $C_D$ reconstruction RMSE against
  $\alpha_{T/4}$, and (\textit{c}) the leave-one-out $R^2$ for each Fourier
  coefficient. The mean drag and $a_2^D$ reach $R^2 = 0.99$, the
  higher harmonics less.}
  \label{fig:appD_err}
\end{figure}

The predictor recovers the mean drag at $R^2 = 0.99$, the second harmonic at $0.99$ and $0.72$, and the fourth-harmonic coefficients at $0.71$ and $0.85$ (figure~\ref{fig:appD_err_c}). The cycle-averaged drag clusters tightly along the diagonal (figure~\ref{fig:appD_err_a}), and over the whole sweep the median RMSE is $0.127$ (figure~\ref{fig:appD_err_b}). The largest errors are at high pitch amplitude, where the drag itself is largest, and in the propulsion corner at high $h_0$ and low $\theta_0$, where the mean drag turns negative and the foil produces net thrust.
%

\section{Moment instantaneous-response validation}\label{sec:appendix_sweep}

\begin{figure}
    \centering
    \begin{subfigure}[t]{0.33\textwidth}\caption{\hfill$(0.50,\,45^\circ)$\hfill}\label{fig:appM_sweep_a}
      \includegraphics[width=\linewidth]{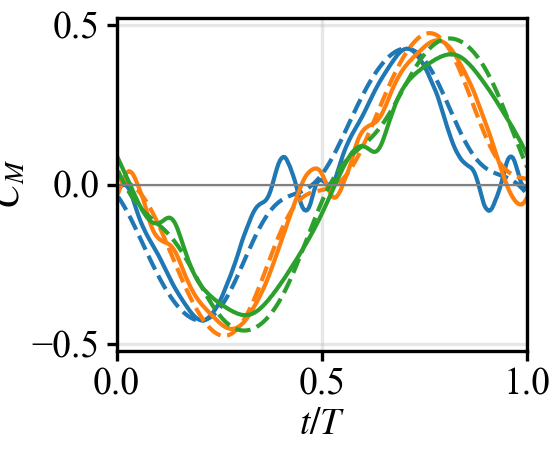}\end{subfigure}\hfill
    \begin{subfigure}[t]{0.33\textwidth}\caption{\hfill$(1.00,\,60^\circ)$\hfill}\label{fig:appM_sweep_e}
      \includegraphics[width=\linewidth]{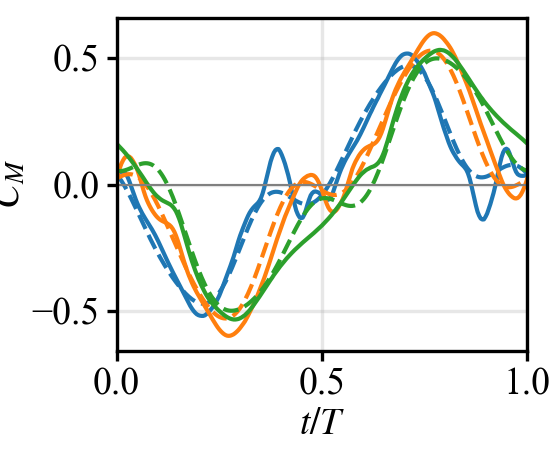}\end{subfigure}\hfill
    \begin{subfigure}[t]{0.33\textwidth}\caption{\hfill$(1.50,\,75^\circ)$\hfill}\label{fig:appM_sweep_i}
      \includegraphics[width=\linewidth]{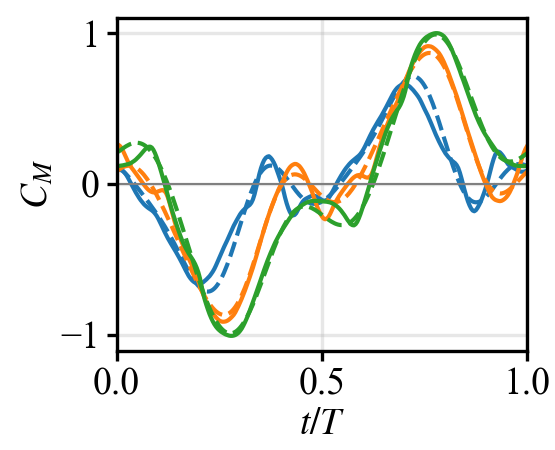}\end{subfigure}\\[3pt]
    \includegraphics[width=0.5\textwidth]{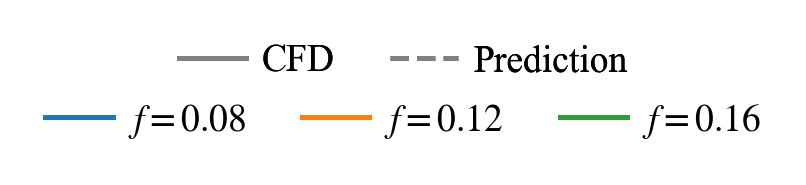}
    \caption{Simulation (solid) versus closed-form prediction (dashed) of $C_M(t)$ across the kinematic sweep. Each panel is labeled by its $(h_0, \theta_0)$.
    Within each panel, three frequencies $f = 0.08, 0.12, 0.16$ are overlaid by color.}
    \label{fig:appM_sweep}
\end{figure}

This appendix collects results from the moment analysis. 
The reconstruction follows the simulation across the sweep at a median RMSE of $0.051$ (figure~\ref{fig:appM_sweep}). The largest value, $0.255$, occurs at $(f, h_0, \theta_0) = (0.16, 1.50, 40^\circ)$, the same combination that gives the largest lift error. Over the energy-harvesting combinations, the peak error has a median of $5.1\%$ and a mean of $6.6\%$, and the phase error has a median of $0.004$, with $99\%$ of the predictions within $0.02$. Both errors are about half the lift values of \S\ref{sec:error}.  The outliers follow the same pattern as the lift, occurring where the mean power approaches zero, with one peak error of $114\%$ and two phase errors above $0.02$.


\bibliographystyle{jfm}
\bibliography{jfm}

\end{document}